\documentclass[11pt]{article}
\RequirePackage[OT1]{fontenc}
\usepackage{amsthm,amsmath,amssymb,amsfonts,bm,enumerate,xcolor,graphicx,natbib,color,url}
\RequirePackage[colorlinks,citecolor=blue,urlcolor=blue]{hyperref}
\usepackage{hyperref}
\usepackage{subcaption} 
\usepackage{textcmds}
\usepackage{booktabs}
\usepackage{multirow}
\usepackage[toc,page]{appendix} 
\usepackage{algorithm,algorithmic}
\usepackage{lipsum}

\usepackage{setspace}
\allowdisplaybreaks

\usepackage{ulem} 

\def\frac#1#2{{\textstyle{#1\over#2}}}

\DeclareSymbolFont{AMSb}{U}{msb}{m}{n}
\DeclareMathSymbol{\Natural}{\mathbin}{AMSb}{"4E}
\DeclareMathSymbol{\Integer}{\mathbin}{AMSb}{"5A}
\DeclareMathSymbol{\Real}{\mathbin}{AMSb}{"52}
\DeclareMathSymbol{\Rational}{\mathbin}{AMSb}{"51}
\DeclareMathSymbol{\Imaginary}{\mathbin}{AMSb}{"49}
\DeclareMathSymbol{\Complex}{\mathbin}{AMSb}{"43} 
\DeclareMathSymbol{\Disk}{\mathbin}{AMSb}{"44} 
\def\bi{\begin{itemize}}
\def\ei{\end{itemize}}
\def\bd{\begin{description}}
\def\ed{\end{description}}
\def\ben{\begin{enumerate}}
\def\een{\end{enumerate}}
\def\bv{\begin{verbatim}}
\def\ev{\end{verbatim}}

\def\hat#1{{\widehat{#1}}}

\def\pr{{\rm Pr}}
\def\Pr{\pr}

\def\2to{{\ {\buildrel 2\over \longrightarrow}\ }}

\def\I1ton{{$I_1,\ldots,I_n$}}
\def\X1ton{{$X_1,\ldots,X_n$}}
\def\Y1ton{{$Y_1,\ldots,Y_n$}}
\def\Z1ton{{$Z_1,\ldots,Z_n$}}
\def\R1ton{{$R_1,\ldots,R_n$}}
\def\e1ton{{$e_1,\ldots,e_n$}}
\def\t1ton{{$t_1,\ldots,t_n$}}
\def\x1ton{{$x_1,\ldots,x_n$}}
\def\y1ton{{$y_1,\ldots,y_n$}}
\def\z1ton{{$z_1,\ldots,z_n$}}

\usepackage[english]{babel}
\usepackage{amsfonts}
\usepackage{amssymb}
\usepackage{graphicx}
\usepackage{enumerate}

\newcommand{\blind}{1}

\begin{document}
\thispagestyle{empty}
\baselineskip=28pt
\vskip 5mm

\begin{center} {\Large{\bf A flexible Bayesian hierarchical  modeling framework for  spatially dependent peaks-over-threshold data}}
\end{center}

\baselineskip=12pt

\vskip 5mm

\if1\blind
{
\begin{center}
\large
Rishikesh Yadav$^1$, Rapha\"el Huser$^1$ and Thomas Opitz$^2$\\ 
\end{center}
\footnotetext[1]{
\baselineskip=10pt Computer, Electrical and Mathematical Sciences and Engineering (CEMSE) Division, King Abdullah University of Science and Technology (KAUST), Thuwal 23955-6900, Saudi Arabia. E-mails: rishikesh.yadav@kaust.edu.sa; raphael.huser@kaust.edu.sa}
\footnotetext[2]{
\baselineskip=10pt  INRAE, UR546 Biostatistics and Spatial Processes, 228, Route de l'A\'erodrome, CS 40509, 84914 Avignon, France. E-mail: thomas.opitz@inra.fr}
} \fi

\baselineskip=26pt
\vskip 4mm
\centerline{\today}
\vskip 6mm

{\large{\bf Abstract}} In this work, we develop a constructive modeling framework for extreme threshold exceedances in repeated observations of spatial fields, based on general product mixtures of random fields possessing light or heavy-tailed margins and various spatial dependence characteristics, which are suitably designed to provide high flexibility in the tail and at sub-asymptotic levels. Our proposed model is akin to a recently proposed Gamma-Gamma model using a ratio of processes with Gamma marginal distributions, but it possesses a higher degree of flexibility in its joint tail structure, capturing strong dependence more easily.  We focus on constructions with the following three product factors, whose different roles ensure their statistical identifiability: a heavy-tailed spatially-dependent field, a lighter-tailed spatially-constant field, and another lighter-tailed spatially-independent field. Thanks to the model's hierarchical formulation, inference may be conveniently performed based on Markov chain Monte Carlo methods. We leverage the Metropolis adjusted Langevin algorithm (MALA) with random block proposals for latent variables, as well as the stochastic gradient Langevin dynamics (SGLD) algorithm for hyperparameters, in order to fit our proposed model very efficiently in relatively high spatio-temporal dimensions, while simultaneously censoring non-threshold exceedances and performing spatial prediction at multiple sites. The censoring mechanism is applied to the spatially independent component, such that only univariate cumulative distribution functions have to be evaluated. We explore the theoretical properties of the novel  model, and illustrate the proposed methodology by simulation and application to daily precipitation data from North-Eastern Spain measured at about $100$ stations over the period 2011--2020.


\baselineskip=26pt




\baselineskip=26pt
\section{Introduction} 
\label{sec:introduction}
Due to its importance in quantifying risk, the statistical modeling of extreme events is crucial in a wide range of environmental applications. Most environmental data are spatial or spatio-temporal in nature, and models at the interface between spatial statistics and   Extreme-Value Theory (EVT) provide a mathematically rigorous way to study the marginal behavior of extreme events, while accounting for their potentially strong spatial or spatio-temporal dependence; see, e.g., the review papers by \citet{Davison.etal:2012}, \citet{cooley2012survey}, \citet{Davison.Huser:2015}, \citet{Davison.etal:2019}. The block maximum (BM) and peaks-over-threshold (POT) approaches are the two principal techniques for modeling the extremes of a
probability distribution. Although their pros and cons are still debated \citep{bucher2021horse}, the POT approach is usually preferred because of its more natural (and often better) use of available data, its direct modeling of the spatial extreme events that effectively took place, and its ability to model clusters of extreme events. The generalized Pareto (GP) distribution plays a key role in the POT approach, being the only possible limit for the marginal distribution of appropriately rescaled high threshold exceedances; see \citet{Davison.Smith:1990}.

Various statistical approaches have been proposed for
the modeling of spatial extremes, including Bayesian hierarchical models, copula models, max-stable random
fields, and generalized Pareto processes; see the review papers by \citet{Davison.etal:2012} and \cite{huser2020advances}. 
Recently, Bayesian hierarchical models have gained in popularity for modeling spatial extremes due to their flexibility to capture complex spatio-temporal trends, and their ease of inference in both BM and POT settings, when the data can be assumed to be conditionally independent given some spatially-structured latent variables. 
 Conditional independence is a common assumption in Bayesian hierarchical models \citep{Cressie:1993,cooley2007bayesian, Banerjee.etal:2014,  opitz2018inla, johannesson2019approximate} and  drastically simplifies computations, but it also poses a risk of misrepresenting the data-level dependence structure. Based on this assumption, several Bayesian hierarchical models using latent variables have been proposed in the literature to model high threshold exceedances; see, e.g., \citet{cooley2007bayesian}, \citet{opitz2018inla}, \citet{bacro2020hierarchical}, and the recent book chapter by \citet{hazrabook2021}. In particular, \citet{cooley2007bayesian} used Gaussian processes to capture latent spatial dependence and trends in precipitation data, and used a relatively simple Markov chain Monte Carlo (MCMC) algorithm for the estimation of posterior distributions. Similarly, \citet{turkman2010asymptotic} fitted Bayesian hierarchical models to spatio-temporal wildfire data from Portugal by taking advantage of MCMC based inference, while \citet{opitz2018inla} and \citet{castro2019spliced} exploited the integrated nested Laplace approximation (INLA) to fit a model to spatio-temporal threshold exceedances. More recently, \citet{hazrabook2021} proposed using Max-and-Smooth, an approximate Bayesian algorithm designed for extended latent Gaussian models and they applied it to a large-scale extreme precipitation dataset.  
 
However, while these hierarchical models typically succeed in estimating marginal distributions accurately, the conditional independence assumption at the data level is  very restrictive when the goal is to estimate return levels of spatial aggregates. Conditional independence models indeed yield unrealistic realizations for spatial phenomena such as rainfall or temperature, where threshold exceedances usually produce smooth surfaces \citep{ribatet2012bayesian}. \cite{sang2010continuous} realized the limitation of the conditional independence model proposed by \cite{sang2009hierarchical} and  improved it by introducing a Gaussian copula at the data-level. Similarly, \cite{clark2021class} proposed spatial models based on self-exciting processes to capture strong spatial dependence, allowing both data-level and latent-level dependencies.
 
As an alternative solution, in this work we extend the hierarchical modeling framework developed by  \cite{yadav2019spatial}  to capture relatively strong spatial dependence among threshold exceedances, by mixing several random fields multiplicatively in a way that ensures a heavy-tailed marginal behavior and generates a wide range of joint tail structures.
   To propose flexible heavy-tailed models, \cite{yadav2019spatial} relied on Breiman's Lemma \citep{Breiman.1965}, which characterizes the tail behavior of the product of two  nonnegative independent random variables when one of them has power-law tail decay. Let $X_1$ and $X_2$ be nonnegative independent random variables such that $\mathbb{E}(X_1^{\alpha+\epsilon})<\infty,$ for some $\epsilon>0$, and the distribution of $X_2$ is regularly varying at $\infty$ with index $-\alpha<0$, i.e., 
$\Pr\left(X_2 >x\right)=\ell(x)x^{-\alpha}$ , where $\ell(x)>0$ and ${{\ell(tx)}/{\ell(t)}}\rightarrow 1$, as $t\to \infty$. Then, we have the expansion 
\begin{align}
\label{eq:BreimanLemma}
\Pr(X_1 X_2>x)\sim \mathbb{E}(X_1^{\alpha})~\Pr\left(X_2 >x\right), \qquad x\to \infty.
\end{align}
Essentially, the result in \eqref{eq:BreimanLemma} implies that the tail decay behavior of the product of two independent random variables, where one is regularly varying and other is lighter-tailed, is completely determined by the tail behavior of the regularly varying component, while the lighter-tailed component only contributes a constant scaling factor of tail probabilities. Breiman's Lemma  \eqref{eq:BreimanLemma} motivates the construction of models with improved flexibility at a sub-asymptotic level, while allowing a heavy-tailed behavior. \cite{yadav2019spatial} used this result to generalize the GP distribution, which can be obtained from the product of two independent Exponential and Inverse Gamma-distributed random variables. Specifically, they proposed  spatial models constructed as $Y(\bm s)= X_1(\bm s) X_2(\bm s)$, where $X_1(\bm s)$ and $X_2(\bm s)$ are independent processes that have Gamma and Inverse Gamma margins, respectively. While the light-tailed Gamma process $X_1(\bm s)$ was assumed to be spatial white noise, the heavy-tailed Inverse Gamma process $X_2(\bm s)$ was used to incorporate spatial dependence, thus inducing dependence among threshold exceedances. 
However, because of the spatial independence of $X_1(\bm s)$, the range of possible dependence structures that the product ${Y(\bm s)=X_1(\bm s) X_2(\bm s)}$ can attain is very limited, preventing the model from capturing strong tail dependence. This issue can also be seen by reformulating the process $Y(\bm s)$ as a Bayesian hierarchical model, whereby $Y(\bm s)\mid X_2(\bm s)$ is a conditionally independent Gamma process and $X_2(\bm s)$ is a latent spatially-structured random field. The conditional independence assumption at the data-level here strongly restricts the form of dependence of $Y(\bm s)$.
In this paper, we generalize the hierarchical spatial model of \cite{yadav2019spatial} to  provide new, more flexible hierarchical spatial models for threshold exceedances, that mitigate the effect of the conditional independence assumption at the data-level with the ultimate goal of capturing stronger spatial dependence among extreme events, while retaining the computational benefits and the intuitive interpretation of such Bayesian hierarchical models. 
For the modeling of spatial precipitation (for which we usually find heavy tails), we consider instead the product of three suitably defined processes possessing different marginal and dependence characteristics, and with clearly distinct roles in the overall spatial model. More precisely, we assume that our model can be written as 
\begin{align}
\label{eq:Prod_process}
Y(\bm s)=\alpha(\bm s) X_1(\bm s) X_2(\bm s) X_3(\bm s),
\end{align}
 where $X_1(\bm s)\geq 0$ is a unit mean noise process with independent and identically distributed (\emph{iid}) variables that captures small scale variations and that allows fast Bayesian computations; $X_2(\bm s)\equiv X_2\geq 0$ is a fully dependent spatial process with unit mean, which counterbalances the noise process $X_1(\bm s)$ in case of strong dependence; $X_3(\bm s)\geq 0$ is a spatial process with unit mean and non-trivial spatial dependence structure, which captures the decay of spatial dependence with respect to spatial distance; $X_1(\bm s)$, $X_2(\bm s)$ and $X_3(\bm s)$ are mutually independent; and $\alpha(\bm s)>0$ is a spatially-varying scale parameter capturing non-stationarity in terms of covariates, where $\alpha(\bm s)=\mathbb{E}[Y(\bm s)]$ owing to the unit mean condition on the random product terms; see Section \ref{subsec:GenModels} for more details. Through their marginal distributions, these three underlying random processes are appropriately weighted to  determine the extent to which they contribute to the overall product mixture. We carefully design the roles of the three components to ensure their identifiability and to allow for meaningful interpretations. Our constructive modeling framework provides flexible models for the upper tail, and sub-asymptotic levels, such that we can use them for modeling complete datasets with heavy-tailed margins and moderate to strong upper tail dependence.  In our models, the strength of tail dependence relies on the choice of the underlying copula in the latent process $X_3(\bm s)$, and this gives flexibility to capture various asymptotic dependence regimes; see Section \ref{subsec:jointtail} for more details.

In this work, we focus on modeling heavy-tailed data, though the model could also be adapted to light tails with exponential decay by switching to an additive structure via a logarithmic transformation of \eqref{eq:Prod_process}, i.e., the data would be represented as $\log\alpha(\bm s) + \log X_1(\bm s) + \log X_2(\bm s) + \log X_3(\bm s)$. Negative values can arise in this construction, but this can be avoided by considering the so-called softplus transformation $\log (1+Y(\bm s))$ instead of $\log Y(\bm s)$, with $Y(\bm s)$ as in \eqref{eq:Prod_process}.

To fit our model to threshold exceedances $Y(\bm s) >u(\bm s)$, where $u(\bm s)$ is some high spatially-varying threshold, we take advantage of the hierarchical formulation of our model, and we exploit customized MCMC methods. Specifically, we use the simulation-based Metropolis adjusted Langevin algorithm (MALA) with random block proposals for latent parameters, as well as the stochastic gradient Langevin dynamics (SGLD) algorithm \citep{welling2011bayesian} for hyperparameters. The SGLD algorithm combines the popular stochastic gradient descent algorithm, known as an important optimization method in Machine Learning \citep{neal2012bayesian,deng2018bayesian, zhang2020cyclical}, and the Langevin dynamics \citep{neal2011mcmc}, to tackle problems in relatively high spatio-temporal dimensions. The SGLD algorithm indeed significantly reduces the computation burden to fit our model to threshold exceedances, and allows inference for massive datasets, by contrast to alternative, more classical inference methods for extreme-value POT models \citep{thibaud2015efficient, de2018high, Huser.Wadsworth:2019}. In our modeling framework, low values such that $Y(\bm s)\leq u(\bm s)$ are completely censored in the MCMC algorithm. This censoring mechanism can be performed very efficiently thanks to the independent noise component $X_1(\bm s)$ in \eqref{eq:Prod_process}; see also \citet{zhang2021hierarchical} for a related censored inference approach. 

The paper is organized as follows. In Section \ref{sec:GenModelsFull}, we define our general product mixture model for spatial extremes with a specific example and study its joint tail behavior. In Section \ref{sec:SimInferenceCens}, we provide censored likelihood expressions for the product mixture model and a simulation study to check the performance of the MCMC sampler that are based on the MALA/SGLD algorithm, and in Section \ref{sec:application}, we fit our model to daily mean precipitation intensities in North-Eastern Spain observed at $94$ monitoring stations. Conclusions and some future research directions are detailed in Section \ref{sec:conclusion}.

\section{Product mixture models for spatial threshold exceedances}
\label{sec:GenModelsFull}
\subsection{General construction}
\label{subsec:GenModels}
Let $Y_t(\bm s)$, $\bm s\in \mathcal{S}\in \mathbb R^2$, be the spatial process of interest observed at time $t \in \{1,
\ldots,n\}$ and at finite set of $d$ locations $\bm s_1,\ldots,\bm s_d\in \mathcal{S}$,
and let the random vector  $\bm Y_t=(Y_{t1},\ldots,Y_{td})^T=\{Y_t(\bm s_1),\ldots,Y_t(\bm s_d)\}^T$ denote the $t^{\text{th}}$ observed time replicate. We assume that the processes $Y_t(\bm s)$, $ t=1,\ldots,n$, are \emph{iid} copies of the process $Y(\bm s)$ in \eqref{eq:Prod_process}, such that $Y_t(\bm s)=\alpha(\bm s) X_{1t}(\bm s) X_{2t}(\bm s) X_{3t}(\bm s)$, for some non-negative independent processes $X_{1t}(\bm s), X_{2t}(\bm s)$ and $X_{3t}(\bm s)$. Similarly, we write $\bm X_{1t}=(X_{1t1},\ldots, X_{1td})^T=\{X_{1t}(\bm s_1),\ldots, X_{1t}(\bm s_d)\}^T$, $\bm X_{2t}=(X_{2t1},\ldots, X_{2td})^T=\{X_{2t}(\bm s_1),\ldots, X_{2t}(\bm s_d)\}^T$, $\bm X_{3t}=(X_{3t1},\ldots, X_{3td})^T=\{X_{3t}(\bm s_1),\ldots, X_{3t}(\bm s_d)\}^T$, and $\bm \alpha=(\alpha_1,\ldots, \alpha_d)^T=\{\alpha(\bm s_1),\ldots,\alpha(\bm s_d)\}^T$. We now describe each of the terms in this product mixture.  We assume that $X_{1t}(\bm s)$ is a noise process such that the variables $X_{1t}(\bm s_j)$ are \emph{iid} with distribution function $F_1$ and mean one, i.e., $\mathbb{E}(X_{1tj})=1$, for $j=1,\ldots,d, t=1\ldots,n$. We further assume that $X_{2t}(\bm s)$ is spatially constant, such that, almost surely, $X_{2t1}=\cdots=X_{2td}\equiv X_{2t} \stackrel{iid}{\sim} F_2$, for some distribution $F_2$, and that $\mathbb{E}(X_{2t})=1,t=1,\ldots,n$.  Specifically, we assume that $F_1$ and $F_2$ have Weibull-like tails, i.e., $1-F_1(x)\sim x^{\eta}\exp\{-(x/\lambda)^\kappa\}$, as $x\to\infty$, for some $\eta\in\Real$, $\lambda>0$ and $\kappa>0$, and similarly for $F_2$. 
Finally, we assume that $X_{3t}(\bm s)$ is a non-trivial spatial process such that $\bm X_{3t}\stackrel{iid}{\sim} F_{\bm X_3}$, where the joint distribution $F_{\bm X_3}$ has an underlying copula $C_{\bm X_3}$  (i.e., multivariate distribution with fixed uniform margins)  and regularly varying marginal distribution, $F_3$, with index $-1/\xi<0$, for some $0<\xi<1$, and $\mathbb{E}(X_{3tj})=1$, $j=1,\ldots,d,\, t=1,\ldots,n$. Thus, we have that $1-F_1(x)=o(1-F_3(x))$ and $1-F_2(x)=o(1-F_3(x))$, as $x\to\infty$. We assume that the three random fields $X_{1t}(\bm s)$, $X_{2t}(\bm s)$, and $X_{3t}(\bm s)$ are mutually independent (also across time $t$). As for the spatially-varying scale parameter $\alpha(\bm s)$, we model it with covariates through a log-link function. Therefore, our general product mixture model is defined at the observed sites as  
\begin{align}
\label{eq:GenProdModel}
\bm Y_t=\bm\alpha \bm X_{1t}\bm X_{2t}\bm X_{3t}, \quad t=1,\ldots,n,
\end{align}
 where $\bm \alpha=\exp(\gamma_0 \bm{1} + \gamma_1 \textbf{Z}_1+\cdots+\gamma_p \textbf{Z}_p )$ is the scale vector, $\textbf{Z}_1, \ldots,\textbf{Z}_d$ are spatial covariates measured at  the $d$ locations, $\gamma_0,\ldots \gamma_p$ are the corresponding regression coefficients, $n$ denotes the total number of  independent time replicates, and operations are done componentwise.  
 By construction, the mean of the observed vector $\bm Y_t$ in \eqref{eq:GenProdModel} is  $\mathbb{E}(\textbf{Y}_t)=\boldsymbol{\alpha}$, and the marginal tail index is $\xi$ (thanks to Breiman's Lemma \eqref{eq:BreimanLemma}). The three random vectors in the product model \eqref{eq:GenProdModel} are designed to have distinct roles and capture different characteristics. Specifically, 
the random vector $\bm X_{1t}$ is composed of \emph{iid} variables, which allows us to capture small-scale variations and to perform fast Bayesian computations in case non-extreme observations are censored.  The term $\bm X_{2t}$ has a fully dependent spatial structure, which counterbalances the \emph{iid} term $\bm X_{1t}$ in case the data have an overall strong spatial dependence. Finally, $\bm X_{3t}$ has a non-trivial spatial dependence structure, which is needed to capture the decay of spatial dependence with respect to distance. By suitably defining their marginal distributions, $F_1$, $F_2$, and $F_3$, respectively, each of these random fields are appropriately weighted in our model with some ``weight" parameters to be estimated from the data. A specific example is given in Section \ref{subsec:ExampleProdModel}.
 
  From \eqref{eq:GenProdModel}, we can rewrite the spatial product mixture model hierarchically as follows:
\begin{align}
\label{eq:GeneralHighericalModel}
Y_{tj}\mid \bm X_{2t}, \bm X_{3t}, \bm{\Theta}_{\bm X_1} &\stackrel{\text{ind}}{\sim} F_{1} ( \cdot/(\alpha_j X_{3 tj} X_{2t}) \,;\bm {\Theta}_{\bm X_1}),\qquad j=1,\ldots,d,\, t=1,\ldots,n; \\
X_{2t}\mid \bm{\Theta}_{\bm X_2} &\stackrel{\text{ind}}{\sim} F_2 (\,\cdot\,;{\bm{\Theta}_{\bm X_2}}), \qquad t=1,\ldots,n; \nonumber\\
\bm X_{3t} \mid \bm{\Theta}_{\bm{X_3}}^{\text{mar}^T},\bm \Theta_{\bm{X_3}}^{\text{dep}^T} &\stackrel{\text{ind}}{\sim}  C_{\bm{X_3}}\left\{F_{3}(\cdot;\bm \Theta_{\bm X_3}^{{\rm mar}}),\ldots,F_{3}(\cdot;\bm \Theta_{\bm X_3}^{{\rm mar}});\bm \Theta_{\bm X_3}^{{\rm dep}} \right\};\nonumber\\ 
\bm{\Theta}&\sim \pi(\bm{\Theta}),\nonumber
\end{align}
where  $\bm{\Theta}=(\bm{\Theta}_{\bm X_1}^T,{\bm {\Theta}}_{\bm X_2}^T,\bm{\Theta}_{\bm X_3}^{\text{mar}^T},\bm \Theta_{\bm X_3}^{\text{dep}^T})^T$ are the unknown model hyperparameters, $\bm{\Theta}_{\bm X_1}$ and $ {\bm {\Theta}}_{\bm X_2} $ denote the hyperparameter vectors for the marginal distributions of the random vectors $\bm X_{1t}$ and $\bm X_{2t}$, respectively, $\bm \Theta_{\bm X_3}=(\bm{\Theta}_{\bm X_3}^{\text{mar}^T}, \bm \Theta_{\bm X_3}^{\text{dep}^T})^T$ contains parameters for the random vector $ \bm X_{3t}$, with $\bm{\Theta}_{\bm X_3}^{\text{mar}^T}$ controlling its marginal distribution and $\bm \Theta_{\bm X_3}^{\text{dep}^T}$ controlling the dependence structure, $\pi (\bm \Theta)$ denotes the prior distribution for the hyperparameter vector $\bm \Theta$, and $\alpha_j$ denotes the $j^{\text{th}}$ parameter in the scale vector $\bm\alpha$.

The hierarchical construction of the proposed product model \eqref{eq:GeneralHighericalModel} suggests using Bayesian inference based on the full data vector $\bm Y=(\bm Y_1^T,\ldots,\bm Y_n^T)^T$, by treating $\bm X_2=(X_{21},\ldots X_{2n})^T$ and $\bm X_3=(\bm X_{31}^T,\ldots,\bm X_{3n}^T)^T$ as latent variables. These two latent vectors,  $\bm X_2$ and $\bm X_3$, are of dimension $n$ and $nd$, respectively. Therefore, there will be a total of $n+nd+|\bm \Theta|$ latent variables and  hyperparameters to make inference for, simultaneously. Let $\pi(\cdot)$ denote a generic (conditional) density, then the joint posterior density of $\bm \Theta$, $\bm X_2$, and $\bm X_3$, denoted by $\pi_{\text{post}}(\bm \Theta,\bm X_2, \bm X_3\mid \bm Y)$, is proportional to $\pi(\bm Y, \bm \Theta,\bm X_2,\bm X_3)=\pi(\bm Y\mid \bm X_2,\bm X_3,\bm \Theta_{\bm X_1}) \pi(\bm X_2\mid \bm \Theta_{\bm X_2})  \pi(\bm X_3\mid \bm \Theta_{\bm X_3}) \pi(\bm \Theta) $, and the posterior density of $\bm \Theta$ is thus obtained after integrating out the latent variables $\bm X_2$ and $\bm X_3$, i.e.,
\begin{align}
\label{eq:PostHyper}
\pi(\bm \Theta \mid \bm Y)=\iint \pi_{\text{post}}(\bm \Theta, \bm X_2, \bm X_3 \mid \bm Y) {\rm{d}}\bm X_2 {\rm{d}} \bm X_3.
\end{align}
The dimension of the integral in \eqref{eq:PostHyper} is very large. We solve this issue in Section \ref{subsec:SGD} by using a customized MCMC algorithm by combining the Metropolis adjusted Langevin algorithm (MALA) with block proposals and the stochastic gradient Langevin dynamics (SGLD), in order to efficiently generate representative  posterior samples of $\bm \Theta$, $\bm X_2$, and $\bm X_3$ from the target posterior distribution.

\subsection{An example of a flexible product mixture model}
\label{subsec:ExampleProdModel}
The general product mixture model formulation in \eqref{eq:GenProdModel} can be used to construct various specific spatial models with a flexible heavy-tailed behavior in the upper tail. Here, we provide one specific example of spatial product mixture model, which we also use in our simulation study in Section \ref{subsec:jointtail}, and in the data application in Section \ref{sec:application}. Let Exp$(1)$ denote the exponential distribution with rate parameter one. Then, a particular product mixture model can be obtained by specifying the terms in \eqref{eq:GenProdModel} as follows:
\begin{align}
X_{1tj} &=E_{1tj}^{\beta_1}/\Gamma(1+\beta_1), \quad E_{1tj}\stackrel{\emph{iid}} \sim \text{Exp}(1), \quad j=1\ldots,d ,\,t=1,\ldots,n,\quad \beta_1>0; \nonumber \\
X_{2t} &=E_{2t}^{\beta_2}/\Gamma(1+\beta_2), \quad E_{2t}\stackrel{\emph{iid}} \sim \text{Exp}(1), \quad t=1,\ldots,n, \quad \beta_2>0, \nonumber 
\end{align}
where $\Gamma(x)=\int_{0}^{\infty} s^{x-1} e^{-s} \rm{d}s$ is the gamma function. The parameters $\beta_1$ and $\beta_2$ affect the marginal distributions of $\bm X_{1t}$ and $\bm X_{2t}$, respectively, and can be interpreted as ``weights'' in the mixture \eqref{eq:GenProdModel}, though they need not sum to one nor be less than one.  As $\beta_1 \to 0$, the independent term $\bm X_{1t}$ indeed becomes degenerate at one, and is thus negligible in the mixture \eqref{eq:GenProdModel}. This holds similarly for the fully dependent term $\bm X_{2t}$, as $\beta_2\to 0$. To ensure tails that are not heavier than exponential and/or to prevent potential numerical issues when computing the gradient of the log-posterior distribution (see Section \ref{sec:SimInferenceCens} and the Supplementary Material), both $\beta_1$ and $\beta_2$ may be restricted to the range to $(0,1]$, though this is not strictly necessary and in the application in Section~\ref{subsec:DataDescription} we allow slightly heavier-tailed distributions by restricting $\beta_1$ and $\beta_2$ to $(0,2]$ instead.
   As for the random vector $\bm X_{3t}$, we define its marginal distribution $F_3$ to be the Inverse Gamma (IG) distribution with scale  $\beta_3-1,$ and shape $ \beta_3$, for some parameter $\beta_3>1$, such that $\mathbb{E}(X_{3tj})=1$, as desired. Moreover, as $\beta_3\to \infty$, the term $\bm X_{3t}$ becomes degenerate at one and does not have any effect on the mixture \eqref{eq:GenProdModel}. Thus, this spatial model is more suitable for moderately heavy-tailed data with $\xi=1/\beta_3>0$ (i.e., $\beta_3<\infty$).
 In summary, the  marginal distributions $F_1$, $F_2$ and $F_3$ in \eqref{eq:GeneralHighericalModel} can be written as
\begin{align}
\label{eq:PrdocutModelExam}
F_1 = \text{Wb} \left\{\cdot; {{1}\over{\beta_1}},{{1}\over{\Gamma(1+\beta_1)}}\right\},
F_2= \text{Wb} \left\{\cdot; {{1}\over{\beta_2}},{{1}\over{\Gamma(1+\beta_2)}}\right\},
F_3= \text{IG}(\cdot; \beta_3, \beta_3-1),
\end{align}
where Wb$(\cdot;\kappa,\lambda)$ denotes the  Weibull distribution with shape $\kappa>0$ and scale $\lambda>0$, and IG$(\cdot;a,b)$ denotes the Inverse Gamma distribution with shape $a>0$ and scale $b>0$.  Furthermore, let the underlying copula $C_{\bm X_3}$ of $\bm X_{3t}$ be the Gaussian copula  with exponential correlation function $\rho(h)=\exp(-h/\rho), h\geq 0,$ and range $\rho>0$. Then, $\bm X_{3t}$ has joint distribution
\begin{align}
\label{eq:ExamGausCopula}
\bm X_{3t} \mid \bm{\Theta}_{\bm{X_3}}^{\text{mar}^T},\bm \Theta_{\bm{X_3}}^{\text{dep}^T} &\sim   \Phi_{\rho}\biggl( \Phi^{-1}\bigl[\text{IG}\{\cdot;\beta_3, \beta_3-1 \}\bigr],\ldots,\Phi^{-1}\bigl[\text{IG}\{\cdot;\beta_3, \beta_3-1\}\bigr] \biggr),
\end{align}
where $\Phi_{\rho}$ is the multivariate Gaussian distribution function with zero mean and correlation matrix $\bm \Sigma(\rho)$ with entries $\bm \Sigma_{i_1,i_2}=\exp(-\|\bm s_{i_1}-\bm s_{i_2}\|/\rho)$, and $\Phi^{-1}$ is the quantile function of the standard Gaussian distribution. For this specific model, the hyperparameter vector is
 $\bm{\Theta}=(\bm{\Theta}_{\bm X_1}^T,{\bm {\Theta}}_{\bm X_2}^T,\bm{\Theta}_{\bm X_3}^{\text{mar}^T},\bm \Theta_{\bm X_3}^{\text{dep}^T})^T$ with
\begin{align*}
\bm{\Theta}_{\bm X_1}=(\beta_1,\gamma_0,\gamma_1,\ldots,\gamma_p )^T; \quad {\bm {\Theta}}_{\bm X_2}=\beta_2, \quad \bm{\Theta}_{\bm X_3}^{\text{mar}}=\beta_3, \quad\text{and} \quad \bm \Theta_{\bm X_3}^{\text{dep}}=\rho.
\end{align*}
The marginal tail index for this model is $\xi=1/\beta_3>0$, as the IG$(\cdot;a,b)$ distribution is regularly varying with index $a$. Therefore, as described, the marginal distribution of $\bm Y$ is heavy-tailed, and the tail heaviness is controlled with the parameter $\beta_3$.

It is of course also possible to consider alternative copula structures for $\bm X_{3t}$. A natural alternative, which produces asymptotic dependence in the data vector $\bm Y$, is the Student's $t$ copula with dispersion matrix $\bm\Sigma(\rho)$ and degrees of freedom $\nu>0$. 
 \subsection{Joint tail behavior}
 \label{subsec:jointtail}
 We here derive the theoretical joint tail behavior of the spatial product mixture model \eqref{eq:GenProdModel} for the case where the latent vector $\bm X_{3t}$  has a regularly varying marginal distribution with positive tail index $\xi>0$ and $\bm X_{1t}$ and $\bm X_{2t}$ are lighter-tailed such that $\mathbb{E}\{(X_{1tj} X_{2tj})^{1/\xi+\varepsilon}\}=\mathbb{E}(X_{1tj}^{1/\xi+\varepsilon})\mathbb{E}(X_{2tj}^{1/\xi+\varepsilon})<\infty$ for some $\varepsilon>0$, which includes  the example in Section \ref{subsec:ExampleProdModel}.
 For convenience, we drop the subscript $t$ in this subsection, so that the spatial product mixture model \eqref{eq:GenProdModel} may be generally written as $\bm Y=\bm \alpha \bm X_1 \bm X_{2} \bm X_{3}$, with $\bm X_1, \bm X_2, \bm X_3$ defined as above.
Furthermore, let the multivariate distribution $F_{\bm X_3}$ of $\bm X_3$ at a finite number of locations be (jointly) regularly varying at $\infty$ \citep{Resnick.1987} such that
\begin{align*}
{{1-F_{\bm X_3}(z \bm x_3)}\over{1-F_{\bm X_3}(z \boldsymbol{1})}} \to V_{\bm X_3}(\bm x_3),\quad \bm x_3>\bm 0, \quad z\to\infty,
\end{align*}
where $\textbf{1}=(1,\ldots,1)^T\in \mathbb{R}^d$ and $V_{\bm X_3}(\bm x_3)$ is some positive limit function that is homogeneous of order $-1/\xi$, i.e., $V_{\bm X_3}(z \bm x_3)=z^{-1/\xi} V_{\bm X_3}(\bm x_3)$ for all $\bm x_3 > \bm 0$ and $z>0$. Then, Theorem 3 of {\color{blue} Fougeres and Mercadier (2012)}  implies the multivariate regular variation of $F_{\bm Y}$, which denotes the multivariate distribution of the data vector $\bm Y$, i.e.,
\begin{align}
\label{eq:AD_AI}
{{1-F_{\bm Y}(z \bm y)}\over{1-F_{\bm Y}(z \boldsymbol{1})}}\to V_{\bm Y}(\bm y)=\int \limits_{0}^{\infty}  \int \limits_{0}^{\infty} \cdots \int \limits_{0}^{\infty} V_{\bm X_3} \{\bm y/ (\bm\alpha x_2 \bm x_1 )\}\left(\prod_{j=1}^{d}{\rm {d}}F_1(x_{1j})\right) {\rm {d}}F_2(x_2), \quad z \to \infty,
\end{align}
where $V_{\bm Y}(\cdot)$ is some positive limit function that is also homogeneous of order $-1/\xi$. Equation \eqref{eq:AD_AI} fully characterizes the extremal dependence of the product mixture $\bm Y$  resulting from \eqref{eq:GenProdModel} in the heavy-tailed case. More, explicitly,
 the bivariate random vector $\bm Y=(Y_1,Y_2)^T$  is asymptotically independent if and only if $ \bm X_3=(X_{31}, X_{32})^T$ is asymptotically independent.  Therefore, the  asymptotic dependence class of the product mixture $\bm Y$ depends on the choice of copula in the latent vector $\bm X_3$. For example, $\bm Y$ is asymptotically independent when we use a Gaussian copula in $\bm X_3$ and asymptotically dependent when we use a Student's $t$ copula with
$\nu > 0$ degrees of freedom instead.  

Let $Y_1 \sim F_{Y_1}$ and $Y_2 \sim F_{Y_2}$, then a summary of the extremal dependence
strength is the tail correlation coefficient $ \chi = \lim_{u\to1}  \chi(u)$, where  $\chi(u) = \Pr\{Y_1 > F_{Y_1}^{-1}(u) \mid Y_2 > F_{Y_2}^{-1}(u)\}$.
\begin{figure}[t!]
\includegraphics[width=1.0\textwidth]{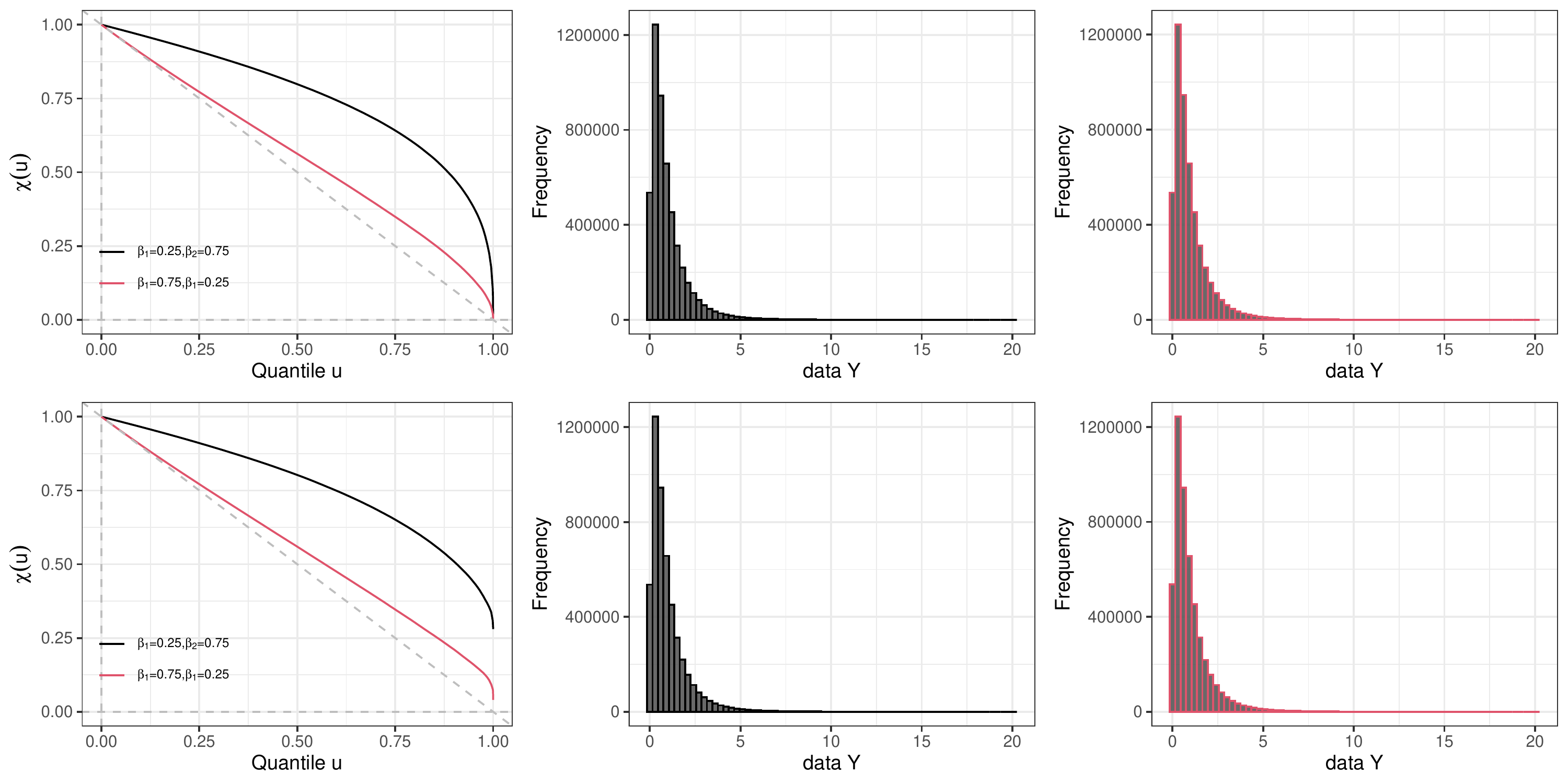}
\caption{ Plot of $\chi(u)$ (first column) for the product mixture model \eqref{eq:GenProdModel} with specific structure detailed in Section \ref{subsec:ExampleProdModel}, based on a latent Gaussian copula (top) or Student's $t$ copula (bottom), with parameters chosen as $(\beta_1,\beta_2)=(0.25,0.75)$ (black) or $(0.75,0.25)$ (red), $\alpha_1=\alpha_2=1$, $\beta_3=5$, $\rho=1$, and for Student's $t$ case, $\nu=1$, as well as the respective marginal histograms (second and third columns).}
\label{fig:chiPlot}
\end{figure}
 We now study the strength of extremal dependence of the product mixture model defined in Section \ref{subsec:ExampleProdModel} in terms of the coefficients $\chi(u)$ and $\chi$ when using a latent Gaussian copula, or a latent Student's $t$ copula with $\nu>0$ degrees of freedom. Figure \ref{fig:chiPlot} shows the plot of $\chi(u)$ as a function of $u$ and its limit $\chi$ (first column), and the marginal histogram (last two-columns) when using a latent Gaussian copula (first row) and latent Student's $t$ copula (second row), with an underlying exponential correlation function with range $\rho=1$, at a spatial distance $0.5$ (i.e., for a correlation of $\exp(-0.5)\approx 0.61$ in the latent vector $\bm X_3$). The other hyperparameters are set as follows: $\alpha_1=\alpha_2=1$, $(\beta_1, \beta_2)^T=(0.25,0.75)^T$ or $(0.75,0.25)^T$, and for the Student's $t$ case, $\nu=1$.  These plots demonstrate that our product model can indeed capture various tail dependence decays, with stronger tail dependence when $\beta_2$ is higher. This is expected as $\beta_2$ is the ``weight'' associated with the fully dependent term $\bm X_2$ in the mixture model described in Section \ref{subsec:ExampleProdModel}. Also, all the histograms appear to be quite similar to each other, which shows that the different parameter combinations give flexibility to capture different types of joint tail behavior, while the marginal tail behavior remains relatively unaffected when the scale parameters $\alpha_1,\alpha_2,$ and the shape parameter $\beta_3$ (i.e., the reciprocal of the tail index) are kept fixed.  Moreover, as expected, the dependence strength of the data vector $\bm Y$ depends on the choice of the copula in the latent vector $\bm X_3$, i.e., the limiting tail correlation coefficient, $\chi$, is strictly positive when we use a latent Student's $t$ copula, whereas  $\chi=0$ when we use a latent Gaussian copula instead.
\section{Simulation-based Bayesian inference}
\label{sec:SimInferenceCens}
\subsection{General strategy}
\label{subsec:GenStatSim}
The hierarchical construction \eqref{eq:GeneralHighericalModel} of our proposed spatial product mixture model \eqref{eq:GenProdModel} naturally suggests using simulation-based Bayesian inference where latent variables are simulated conditional to observations. We use Markov chain Monte Carlo (MCMC) methods to simultaneously generate samples of the hyperparameter vector $\bm \Theta$ and the two latent parameter vectors, $\bm X_2$ and $\bm X_3$. As we fit our model to threshold exceedances,  we describe in Section \ref{subsec:CensLik} the censoring mechanism focusing on the specific model described in Section \ref{subsec:ExampleProdModel}. In Section \ref{subsec:SGD}, we detail our MCMC sampler combining block Metropolis adjusted Langevin algorithm (MALA) updates for latent variables and the stochastic gradient Langevin dynamics (SGLD) for hyperparameters. We demonstrate the performance of our MCMC sampler based on a simulation study in Section \ref{subsec:simstudy}. 

\subsection{Censored likelihood with latent variables, priors, and posterior density}
\label{subsec:CensLik}
We follow the notation of Section \ref{subsec:GenModels}, with lowercase letters denoting realized values, i.e., $y_{tj}$ is the realization of $Y_{tj}=Y_t(\bm s_j)$, $x_{1tj}$ is the realization of $X_{1tj}=X_{1t}(\bm s_j)$, and so forth. Let $\bm e_t=(e_{t1},\ldots,e_{td})^T$ be the exceedance indicator vector, such that $e_{tj}=1$, if $y_{tj}>u_{tj}$, and $e_{tj}=0$, if $y_{tj}\leq u_{tj}$, where $\bm u_t=(u_{t1,}\ldots,u_{td})^T$ is a fixed threshold vector.  If $u_{tj}=0$, no censoring is applied to the value $y_{tj}$, whereas if $u_{tj}=\infty$, the observation $y_{tj}$ is treated as fully censored and as a variable to predict.  Then, the augmented censored likelihood contribution for the parameter $(\bm \Theta^T, x_{2t}, \bm x_{3t}^T)^T$, based on the data vector $(\bm y_t^T, \bm e_t^T)^T$, which stems from the general product mixture model \eqref{eq:GenProdModel} with an underlying copula $C_{\bm X_3}$ (with density $c_{\bm X_3}$), is
\begin{align}
\label{eq:censoredlik}
L(\bm\Theta, x_{2t}, \bm x_{3t} ; \bm y_t, \bm e_t) &=\prod_{j=1}^{d} \left\{ {{1}\over{\alpha_j  x_{2t} x_{3t}}} f_1\left( {{y_{tj}}\over{\alpha_j x_{2t} x_{3tj} }}; \bm \Theta_{\bm X_1}\right)\right\}^ {\mathbb{I}(e_{tj}=1)} \nonumber\\
& \times \prod_{j=1}^{d} \left\{F_1\left({{u_{tj}}\over{\alpha_j x_{2t} x_{3tj}}}; \bm \Theta_{\bm X_1}\right)\right\}^{\mathbb{I}(e_{tj}=0)} \nonumber  \\&
  \times f_2(x_{2t}; \bm\Theta_{\bm X_2})\nonumber
  \\ & 
  \times  c_{\bm{X_3}}\left\{F_{3}(x_{3t1};\bm \Theta_{\bm X_3}^{{\rm mar}}),\ldots,F_{3}(x_{3td};\bm \Theta_{\bm X_3}^{{\rm mar}});\bm \Theta_{\bm X_3}^{{\rm dep}} \right\} \prod_{j=1}^d f_3(x_{3tj};\bm \Theta_{\bm X_3}^{\text{mar}}) ,
\end{align}
where  $f_1$, $f_2$ and $f_3$ are the density functions corresponding to the cumulative  distribution function $F_1$, $F_2$ and $F_3$, respectively, and $\mathbb{I}(\cdot)$ is the indicator function. 
 The expressions in \eqref{eq:censoredlik} correspond to the likelihood contribution of non-censored observations (first line) and censored observations (second line), to the likelihood contribution of the latent variable $ x_{2t}$ (third line) and the latent variable vector $\bm x_{3t}$ (fourth line).
 In particular, when $C_{\bm X_3}$ is the Gaussian copula, the density $c_{\bm X_3}$ has the form 
\begin{align*}
\phi_{\rho}\left\{\Phi^{-1}(v_1),\ldots,\Phi^{-1}(v_d)\right\} \left[\prod_{j=1}^d \phi\{\Phi^{-1}(v_j)\}\right]^{-1}, \quad \bm v=(v_1,\ldots,v_d)^T\in [0,1]^d,
\end{align*}
where $\phi_{\rho}$ and $\phi$ are the multivariate and univariate Gaussian densities corresponding to $\Phi_{\rho}$ and $\Phi$ introduced in \eqref{eq:ExamGausCopula}. Assuming independent time replicates, the overall augmented censored likelihood is then
\begin{align}
\label{eq:OverallCenLik}
L(\bm\Theta, \bm x_{2}, \bm x_{3} ; \bm y, \bm e)=\prod_{t=1}^{n} L(\bm\Theta,  x_{2t}, \bm x_{3t} ; \bm y_t, \bm e_t),
\end{align}
where $\bm y= (\bm y_1^T,\ldots,\bm y_n^T)^T$ is the complete data vector, $\bm x_2 = (x_{21},\ldots,x_{2n})^T$, $\bm x_3=(\bm x_{31}^T,\ldots,\bm x_{3n}^T)^T$, and $\bm e=(\bm e_1^T,\ldots,\bm e_n^T)^T$  is the complete threshold indicator vector. In particular, the overall augmented likelihood for the specific model in Section \ref{subsec:ExampleProdModel} may be obtained by using the distribution functions $F_1$, $F_2$, $F_3$, and their corresponding density functions as specified in \eqref{eq:PrdocutModelExam}. Using \eqref{eq:OverallCenLik}, the joint posterior of model parameters $(\bm \Theta^T, \bm x_2^T, \bm x_3^T)^T$ may be written as 
\begin{align}
\label{eq:Jointpost}
\pi_{\text{post}}(\bm \Theta, \bm x_2, \bm x_3 \mid \bm y, \bm e ) \propto L(\bm\Theta, \bm x_{2}, \bm x_{3} ; \bm y, \bm e) \pi(\bm \Theta).
\end{align}

For convenience, we assume independent vague prior distributions for the model hyperparameters. In particular, for the specific model in Section \ref{subsec:ExampleProdModel}, we have $\pi(\bm \Theta)=\pi(\gamma_0)\times \pi(\gamma_1)\times \cdots \times \pi(\gamma_p)\times \pi(\beta_1)\times\pi(\beta_2)\times \pi(\beta_3)\times \pi(\rho)$. In the application in Section~\ref{subsec:DataDescription}, for instance, we choose a standard Gaussian distribution with mean  $0$ and variance $100$ for all the covariate coefficients  $\gamma_0,\gamma_1,\ldots,\gamma_p$ controlling the scale vector $\bm\alpha=\exp(\gamma_0 \textbf{1} + \gamma_1 \textbf{Z}_1+\cdots+\gamma_p \textbf{Z}_p )$, a uniform distribution on $[0,2]$ for the ``weight'' parameters $\beta_1$ and $\beta_2$ to prevent overly large values, a gamma distribution with shape $1/3$ and rate $1/100$ for $\beta_3$ (reciprocal tail index), and a uniform distribution on $[0,2\delta]$ for $\rho$ (range parameter), where $\delta$ is the maximum spatial distance between the stations. 
\label{sec:augment}
\subsection{ Stochastic gradient Langevin dynamics (SGLD) algorithm}
	\label{subsec:SGD}
	When writing the model hierarchically, there are two latent parameter vectors, $\bm x_2$ and $\bm x_3$, of dimensions $n$ and $nd$, respectively. MCMC computations are therefore quite involved, and MCMC chains might mix poorly when $n$ and $d$ are both large. 
	To avoid this issue when the latent parameter vector is of very high dimension, a suitable choice of proposal distributions is required. The Langevin dynamics (\citet{Roberts.Tweedie:1996}, \citet{roberts1998optimal}) provides an efficient way to define proposal distributions, by exploiting information from the gradient of the log-posterior density, and it generally works quite efficiently in reasonably large dimensions.  Let $\pi(\bm z \mid \bm y_n)$ be an arbitrary target posterior density of $m$ components $\bm z=(z_1,\ldots,z_m)^T$ conditioned on $n$ replicated observations $\bm y_n=(y_1,\ldots,y_n)^T$. Then, the proposal distribution based on the Langevin dynamics require to fix a step size parameter $\tau>0$, and to sample proposals $\bm z^p$ from
\begin{align}
\label{eq:mala}
\bm z^{p}\mid \bm z \sim \mathcal{N}\left(\bm z + {{\tau}\over{2}}P\nabla_{\bm z} \log \pi(\bm z\mid\bm y_n), \,\,\tau P\right),
\end{align}
where $\nabla_{\bm z}$ is the gradient operator with respect to the variable $\bm z$ and $\tau P$ is a covariance matrix; see \citet{atchade2006adaptive}. This algorithm is commonly called the Metropolis adjusted Langevin algorithm, MALA in short. 
If the dimension $m$ is high, MALA proposals are feasible only if the gradient of the log-posterior density with respect to $\bm z$ is available in closed form and can be computed efficiently. In our case, we have closed-form expressions of the log-posterior density with respect to both latent variable vectors $\bm x_2$ and $\bm x_3$, and some hyperparameters, $\gamma_0,\gamma_1,\ldots,\gamma_p$, $\beta_1$ and $\beta_2$. For detailed calculations, see Sections 1.2 and 1.3 of the Supplementary materiel. One disadvantage of the classical MALA proposals, however, is that we need to compute the gradient of the log-posterior density at each MCMC iteration, which may be computationally prohibitive when the data dimension is large. To speed up computations, we instead rely on the SGLD algorithm. The latter may be significantly faster than the MALA, which uses the whole dataset at once \citep[as, e.g., in][]{yadav2019spatial}. The stochastic gradient descent algorithm was popularized in Machine Learning for fitting complex Neural Network structures. Similarly, it is possible to exploit simulation-based MCMC inference based on the SGLD algorithm (a combination of stochastic gradient descent algorithm, and the Langevin dynamics), in order to speed up computations significantly.  This method requires to select a batch size $b\in \{1,\ldots,m\}$ and perform the updates based on a randomly selected sub-dataset of size $b$ instead of the full dataset; for more details see \cite{nemeth2020stochastic}, and the pseudo-code in Algorithm \ref{alg:Alg1}.  
\begin{algorithm}[t!]
\begin{algorithmic}[1]
\STATE \textit{Notation}:  $P\setminus L$:  set difference from $P$ to $L$; $|A |$ cardinality of the set $A$; $p$: proposed states; $c:$ current state
\STATE \textit{Input:} $\bm y_{[A_{n},A_{d}]}$: data matrix of spatial dimension $|A_d|$ and temporal dimension $|A_n|$ where $A_d=\{1,\ldots,d\}$ and $A_n=\{1,\ldots, n\}$; $\bm \Theta_k^{0}$, $\bm x_{2_{[A_n]}}^0$, $\bm x_{3_{[A_n, A_d]}}^0$: initial values for hyperparameter vector $\bm \Theta_k$, and latent parameter vectors $\bm x_2$ and $\bm x_3$, respectively;  $b$:  batch size for the SGLD algorithm;  $N_{bn}$: total burn-in samples (for simplicity, we here consider a single burn-in period); adapt: number of iterations after which we adapt the tuning parameters; $S_{c}$=\{$(c+0)\times$adapt, $(c+6)\times$adapt, $(c+12)\times$adapt, \ldots\}, $c=1,2,3,4,5,6$; $N_{m}$: number of MCMC iterations after which we apply a Metropolis--Hastings correction; $N_{h}(=4$): number of blocks for hyperparameter vector $\bm \Theta$; $N_{tot}=N \times N_{m}$: total number of MCMC iterations
\STATE Start with $\bm \Theta_k^{c'}=\bm \Theta_k^0$, $\bm \Theta_k^c=\bm \Theta_k^0$, $\bm x_{2_{[A_n]}}^c=\bm x_{2_{[A_n]}}^0$, and $\bm x_{3_{[A_n, A_d]}}^c=\bm x_{3_{[A_n, A_d]}}^0$
\FOR{$i=1$ to $N$}
\FOR{$j=1$ to $N_{m}$}
\STATE Sample random indices of length $b$ from $A_n$  without replacement, say the resulting set is $A_b=\{a_1,\ldots,a_b\}\subset A_n$
\FOR{$k=1$ to $N_{h}$} 
\STATE \textit{SGLD for $\bm \Theta_k$}: Propose (transformed) hyperparameter vector $\bm\Theta_k^p$ using SGLD \eqref{eq:SGLD} with batch size $b$, i.e., based on observations $\bm y_{[A_b, A_d]}$, and set $\bm \Theta_k^c=\bm \Theta_k^p$ with probability one
\ENDFOR
 \STATE \textit{Block MALA for $\bm x_2$:} Propose (log-transformed) $\bm x_{2_{[A_b]}}^p$ using SGLD \eqref{eq:SGLD} and calculate the acceptance ratio $r_{\bm x_2}=  R_{\bm x_2}(\bm x_{2_{[A_b]}}^p,\bm x_{2_{[A_b]}}^c \mid \bm y_{[A_b, A_d]}, \bm e_{[A_b, A_d]}, \bm \Theta^{c'},\bm x_{3_{[A_b, A_d]}}^c )$ using \eqref{eq:rat_2}
\STATE Generate $U\sim U[0,1]$, \textbf{if} ($U<r_{\bm x_2})$ $\{\bm  x_{2_{[A_n]}}^c= \bm x_{2_{[A_n\setminus A_b]}}^c \cup\, \bm x_{2_{[A_b]}}^p\}$ \textbf{else} $\{\bm x_{2_{[A_n]}}^c=\bm x_{2_{[A_n]}}^c\}$
\STATE \textbf{if} (($i \times N_{m}< N_{bn}$ ) $\&$ $(i \times N_{m}\in S_5)$) $\{$use adaptive strategy to change the  step size parameter$\}$ \textbf{else} $\{\text{no change} \}$
\STATE \textit{Block MALA for $\bm x_3$:}  Propose (log-transformed) $\bm x_{3_{[A_b, A_d]}}^p$ using SGLD \eqref{eq:SGLD} and calculate the acceptance ratio $r_{\bm x_3}=R_{\bm x_3}(\bm x_{3_{[A_b, A_d]}}^p, \bm x_{3_{[A_b, A_d]}}^c \mid \bm y_{[A_b, A_d]}, \bm e_{[A_b, A_d]}, \bm x_{2_{[A_b]}}^c, \bm \Theta^{c'} ) $ using \eqref{eq:rat_3}
\STATE Generate $U\sim U[0,1]$, \textbf{if} ($U<r_{\bm x_3})$ $\{\bm  x_{3_{[A_n, A_d]}}^c= \bm x_{3_{[A_n\setminus A_b, A_d]}}^c \cup \,\bm x_{3_{[A_b, A_d]}}^p\}$ \textbf{else} $\{\bm x_{3_{[A_n,A_d]}}^c=\bm x_{3_{[A_n,A_d]}}^c\}$
\STATE \textbf{if} (($i \times N_{m}< N_{bn}$) $\&$ $(i \times N_{m}\in S_6)$) $\{$use adaptive strategy to change the step size parameter$\}$ \textbf{else} $\{\text{no change}\}$
\ENDFOR
\FOR{$k=1$ to  $N_{h}$}
\STATE \textit{Metropolis--Hastings correction for} $\bm \Theta_k$ : 
Calculate the acceptance ratio $r_{\bm \Theta_k}=R_{\bm \Theta_k}(\bm\Theta_k^c,\bm\Theta_k^{c'} \mid \bm y_{[A_n, A_d]}, \bm e_{[A_n, A_d]}, \bm x_{2_{[A_n]}}^c,\bm x_{3_{[A_n,A_d]}}^c)$ using \eqref{eq:rat_1} 
\STATE Generate $U\sim U[0,1]$, \textbf{if} ($U<r_{\bm \Theta_k}$)  $\{\bm  \Theta_k^{c'} =\bm \Theta_k^c,\,\bm \Theta_k^c=\Theta_k^c\}$ \textbf{else} $\{\bm \Theta_k^{c'}=\bm \Theta_k^{c'},\, \bm \Theta_k^c=\bm\Theta_k^{c'}\}$ 
\STATE \textbf{if} (($i \times N_{m}< N_{bn}$) $\&$ $(i \times N_{m}\in S_k)$) $\{$use adaptive strategy to change  the  step size parameter$\}$ \textbf{else} $\{\text{no change} \}$
\ENDFOR
\ENDFOR
\end{algorithmic}
\caption{Pseudo-code for the SGLD algorithm with Metropolis--Hastings correction}
\label{alg:Alg1}
\end{algorithm}
Intuitively, if $b\ll m$, then computations at each iteration will be much faster. More specifically, let $P=I_m$, then the proposal distribution \eqref{eq:mala} based on sub-dataset of size $b$ may be written as
\begin{align}
\label{eq:SGLD}
\bm z^{p}\mid \bm z \sim \mathcal{N}\left(\bm z + {{\tau n}\over{2b}}\nabla_{\bm z}^{\star}  \log \pi(\bm z\mid \bm y_b), \,\,\tau \right),
\end{align}
where $\bm y_{b}$ is the vector of length $b$ sampled without replacement from the full data vector $\bm y_n$, and ${n\over b}\nabla_{\bm z}^{\star} \log \pi(\bm z \mid \bm y_b)$ is an unbiased estimator of $\nabla_{\bm z} \log \pi(\bm z\mid \bm y_n)$, based on $\bm y_b$ only (i.e., it is computed from the summands in the gradient corresponding to $\bm y_b$, rescaled by $n/b$). It can be shown that the SGLD algorithm \eqref{eq:SGLD} has theoretical guarantees to converge to the exact stationary distribution as the step size $\tau\equiv\tau_s \to 0$, such that $\sum \tau_s=\infty$ and $\sum \tau_s^2 < \infty$, where $\tau_s$ denote the step size at $s^{\text{th}}$ iteration; see \cite{welling2011bayesian} for more details. In practice, it is difficult to choose the optimal $\tau_s$ parameter as it relies on a bias variance trade-off, and is often chosen by cross-validation in the machine learning literature. As it is difficult to choose $\tau_s$ optimally in our complex models, we instead rely on the Metropolis--Hastings corrections applied after a fixed number of iterations. Here, we propose two different SGLD-based MCMC algorithms (i.e., Algorithm \ref{alg:Alg1} and a simpler version thereof, Algorithm 2, described in the Supplementary Material) and we study their performance in our case by simulation in Section \ref{subsec:simstudy} and in the Supplementary Material. Note that, when the SGLD algorithm \eqref{eq:SGLD} is applied to the vector of latent variables in our model, the batch size $b$ will be the same as the number of selected latent variables (involved in the selected batch), thus $n/b=1$ in \eqref{eq:SGLD}, and after Metropolis--Hastings corrections we may interpret it as a block MALA sampling scheme. 

For conciseness, we here report only the details of Algorithm~\ref{alg:Alg1}, which is based on the SGLD algorithm with Metropolis--Hastings corrections, as it turns out to be the best among the two proposed algorithms in terms of  accuracy and computational cost. For details about the Algorithm 2, see Section 2 of the Supplementary Material.

Algorithm \ref{alg:Alg1} is based on the SGLD algorithm and consists of three successive steps. Let the whole hyperparameter vector $\bm \Theta$ be divided in four blocks as $\bm \Theta=(\bm\Theta_1^T, \bm\Theta_2^T,\bm\Theta_3^T, \bm\Theta_4^T)^T$, where $\bm\Theta_1 = (\gamma_0,\gamma_1,\ldots, \gamma_p)^T$, $\bm\Theta_2=\beta_1$, $\bm\Theta_3=\beta_2$, and $\bm\Theta_4=(\beta_3, \rho)^T$.  In the first step, we update the hyperparameters $\bm \Theta_1$, $\bm \Theta_2$, $\bm \Theta_3$, and $\bm \Theta_4$ in four different blocks using the SGLD algorithm \eqref{eq:SGLD} with a random sub-dataset of size $b$. At every MCMC iteration, we propose a new state using the SGLD algorithm \eqref{eq:SGLD} and accept it with probability one,  and then at the end of every fixed number ($N_m$) of iterations we either accept or reject the whole trajectory using the standard Metropolis--Hastings criterion;  see the pseudo-code in  Algorithm \ref{alg:Alg1} for more details. 
 In the second and third steps, we update a random subset of size $b$ of the latent parameters $\bm x_2$ and $\bm x_3$, respectively, through SGLD proposal \eqref{eq:SGLD}, which is the same as  using exact gradient in \eqref{eq:mala}, and after applying the Metropolis--Hastings correction it becomes a block MALA algorithm; see the pseudo-code in  Algorithm \ref{alg:Alg1} for more details.  Denote $q_1(\bm\Theta_k^{p}\mid \bm\Theta_k )$,
 $q_2(\bm x_2^{p} \mid \bm x_2)$, and  $q_3(\bm x_3^{p} \mid \bm x_3)$  the proposal distributions for the hyperparameter vector $\bm \Theta_k$, the latent parameter vector $\bm x_2$, and the other latent parameter vector $\bm x_3$, respectively, where the superscript $p$ refers to proposal values. Let $\alpha_{\bm \Theta_k}(\bm\Theta_k^{p},\bm\Theta_k )$, $\alpha_{\bm x_2}(\bm x_2^{p},\bm x_2)$, and $\alpha_{\bm x_3}(\bm x_3^{p}, \bm x_3)$ denote  the acceptance probabilities for $\bm \Theta_k$, $\bm x_2$, and $\bm x_3$, respectively. Then,
$\alpha_{\bm \Theta_k}(\bm\Theta_k^{p},\bm\Theta_k)=\min\left\{1,R_{\bm \Theta_k}(\bm\Theta_k^{p},\bm\Theta_k \mid \bm y, \bm e, \bm x_2,\bm x_3 )\right\}$, $\alpha_{\bm x_2}(\bm x_2^{p},\bm x_2) =\min\left\{1,R_{\bm x_2}(\bm x_2^{p},\bm x_2 \mid \bm y, \bm e, \bm \Theta ,\bm x_3 )\right\}$, $\alpha_{\bm x_3}(\bm x_3^{p}, \bm x_3) =\min\left\{1,R_{\bm x_3}(\bm x_3^{p}, \bm x_3 \mid \bm y, \bm e, \bm x_2, \bm \Theta )\right\},$ where the corresponding acceptance ratios are 
\begin{align}
  R_{\bm \Theta_k}(\bm\Theta_k^{p},\bm\Theta_k \mid \bm y, \bm e, \bm x_2,\bm x_3 ) &={{L\left\{(\bm\Theta_k^{p},\bm\Theta_{-k}),\bm x_2,\bm x_3; \bm y, \bm e)\right\} \pi(\bm \Theta_k^{p})\, q_1(\bm \Theta_k\mid \bm \Theta_k^{p}) }\over{L\{(\bm\Theta_k,\bm\Theta_{-k}),\bm x_2,\bm x_3; \bm y, \bm e\} \pi(\bm \Theta_k)  \, q_1(\bm \Theta_k^{p}\mid \bm \Theta_k)}}, \,\, k=1,2,3,4, \label{eq:rat_1}  \\  R_{\bm x_2}(\bm x_2^{p},\bm x_2 \mid \bm y, \bm e, \bm \Theta ,\bm x_3 ) &={{L(\bm\Theta,\bm x_2^{p},\bm x_3; \bm y, \bm e) \, q_2(\bm x_2\mid\bm x_2^{p}) }\over{L(\bm\Theta,\bm x_2,\bm x_3; \bm y, \bm e)  \, q_2(\bm x_2^{p}\mid\bm x_2)}},
   \label{eq:rat_2} \\  
R_{\bm x_3}(\bm x_3^{p}, \bm x_3 \mid \bm y, \bm e, \bm x_2, \bm \Theta ) &={{L(\bm\Theta,\bm x_2,\bm x_3^{p}; \bm y, \bm e) \, q_3(\bm x_3\mid \bm x_3^{p}) }\over{L(\bm\Theta,\bm x_2,\bm x_3; \bm y, \bm e)  \, q_3(\bm x_3^{p}\mid \bm x_3)}},
\label{eq:rat_3}
\end{align}
with $L(\bm \Theta, \bm x_2,\bm x_3; \bm y,\bm e)$ the censored likelihood in \eqref{eq:censoredlik}, and $\bm \Theta_{-k}$ denotes the hyperparameter vector after removing the $k^{\text{th}}$ block from the full hyperaprameter vector $\bm \Theta$. 
The step size parameter $\tau$ in \eqref{eq:SGLD} determines the performance of the MCMC sampler and is directly responsible for jump sizes in the chains of $\{\bm \Theta_k, k=1,2,3,4\}$, $\bm x_2$ and $\bm x_3$ at each MCMC iteration. Here, we have in fact six step size parameters to set, namely, $\{\tau_{\bm \Theta_k}, k=1,2,3,4\}$, $\tau_{\bm x_2}$ and $\tau_{\bm x_3}$, corresponding to the SGLD updates for $\bm \Theta_ks$, $\bm x_2$ and $\bm x_3$, respectively. We select them adaptively, throughout the MCMC algorithm, in order to achieve a desired acceptance probability. Let $\tau$ denote a generic step size parameter. We tune  $\tau$ using the adaptive algorithm from \citet{yadav2019spatial}: during an initial burn-in phase  we modify the current value, $\tau_{\text{cur}} $, of $\tau$ every 500 iterations as  $
\tau_{\text{cur}}\mapsto\tau_{\text{new}}:=\exp\left\{(P_{\text{cur}}-P_{\text{tar}})/\theta\right\}\tau_{\text{cur}},
$  where $P_{\text{tar}}$ is the target acceptance probability, 
  $P_{\text{cur}}$ is the current acceptance probability (computed from the last 500 iterations), and $\theta>0$ is a parameter fixed to modulate the rate of change of $\tau$. Here, we set    $P_{\text{tar}}=0.23$ for random walk proposals and $P_{\text{tar}}=0.57$ for SGLD based proposals. In a second burn-in phase, we update the tuning parameters using  the same adaptive strategy only if the acceptance probability drops out of the intervals $[0.15, 0.30]$ and $[0.50,0.65]$ for random walk and SGLD based proposals, respectively. 

Since all  proposals are based on the  Gaussian distribution,
 we transform  the parameters so that their support  becomes the whole real line through the following reparametrization:
\begin{align*}
\tilde{\gamma_l}=\gamma_l,\,l=0,1,\ldots,p, \,\tilde{\beta}_1=\log\left({{\beta_1}\over{\delta_1-\beta_1}}\right), \, \tilde{\beta}_2=\log\left({{\beta_2}\over{\delta_2-\beta_2}}\right),  \, \tilde{\beta}_3=-\log(\beta_3-1),\,\tilde{\rho}=\log\left({{\rho}\over{2\delta-\rho}}\right).
\end{align*}
where $\delta_1$ and $\delta_2$ are the upper limits for the $\beta_1$ and $\beta_2$ parameters, respectively, often set to some finite positive values (e.g., $\delta_1=\delta_2=1$ or $2$) to avoid numerical issues, and $\delta$ is the maximum distance between the stations (i.e., the ``diameter'' of the study region).  The corresponding reverse transformation is 
\begin{align*}
\gamma_l=\tilde{\gamma}_l,\,l=0,1,\ldots,p,  \,\, \tilde{\beta}_1={{\delta_1\exp({\tilde{\beta}}_1)}\over{1+\exp({\tilde{\beta}}_1)}},  \,\, \tilde{\beta}_2={{\delta_2\exp({\tilde{\beta}}_2)}\over{1+\exp({\tilde{\beta}}_2)}},\,\,  \beta_3=1+\exp(-\tilde{\beta}_3), \,\, \rho= {{2\delta\exp({\tilde{\rho}})}\over{1+\exp({\tilde{\rho}})}}.
\end{align*}
The Jacobian matrix of the transformation is $\log(J)=\log(\delta_1)+\tilde{\beta}_1-2\log\{1+\exp({\tilde{\beta}}_1)\}+\log(\delta_2)+\tilde{\beta}_2-2\log\{1+\exp({\tilde{\beta}}_2)\}-\tilde{\beta}_3+ \log(2)+\log(\delta)+\tilde{\rho}-2\log\{1+\exp({\tilde{\rho}})\}$. 
Similarly, we log-transform the latent parameters as $\tilde{\bm x}_2=\log(\bm x_2)$ and $\tilde{\bm x_3}=\log(\bm x_3)$. 

\subsection{Simulation study}
\label{subsec:simstudy}
We now check the performance of our SGLD-based MCMC sampler (Alghorithm \ref{alg:Alg1}) by simulation with respect to different batch sizes $b$. We also provide a simulation study for the comparison between the Algorithm \ref{alg:Alg1} and Algorithm 2; for conciseness, the results are reported in the Supplementary material. We simulate data from the spatial product mixture model specified in Section \ref{subsec:ExampleProdModel} for $d=100$ spatial locations and $n =200$ time replicates (total $100\times 200+ 200=20,200$ latent variables). The spatial locations are generated uniformly in the unit square $[0,1]^2$ and the latent term $\bm X_{3t}$ has an underlying Gaussian copula with a stationary isotropic exponential correlation function $\sigma(h)=\exp(-\|h\|/\rho),$ where $\rho>0$ is the range parameter, set here to $\rho=0.5$.  For the purpose of spatial prediction, we completely mask the data at $20$ sites, such that their simulated values are treated as fully missing. The censoring threshold $u_{tj}$ is here fixed to the site-wise $75\%$ quantile based on the observations $(y_{1j},\ldots,y_{nj})^T$, and  we set  $u_{tj}=\infty$, whenever data $y_{tj}$ is missing.  The parameters are chosen as $\beta_1=0.8$, $\beta_2=0.7$, $\beta_3=5$ (i.e., $\xi=0.2$), and the scale parameter is modeled spatially as 
$\bm\alpha=\exp\bigl(\gamma_0 \bm 1+\gamma_1 \bm Z_1+\gamma_2\bm Z_2+\gamma_3 \bm Z_3\bigr)$, where $\exp(\gamma_0)=\gamma_1=\gamma_2=\gamma_3=1$,  $\bm Z_1$ denotes the $x$-coordinate of each site,  $\bm Z_2$ denotes the $y$-coordinate of each site, and $\bm Z_3$ is a covariate that is randomly generated from the standard normal distribution. 

To make inference, we use the MCMC sampler detailed in Algorithm \ref{alg:Alg1}, where we set $N_m=25$, $\delta_1=\delta_2=1$, so that the parameters $\beta_1$ and $\beta_2$ lie within the interval $[0,1]$, and $\delta$ is set to the maximum spatial distance between the observed locations. We run two MCMC chains in parallel with two different initial values to check the convergence of Markov chains, and we compute the posterior summaries by averaging values from the two Markov chains.
\begin{figure}[t!]
\includegraphics[width=1.0\textwidth]{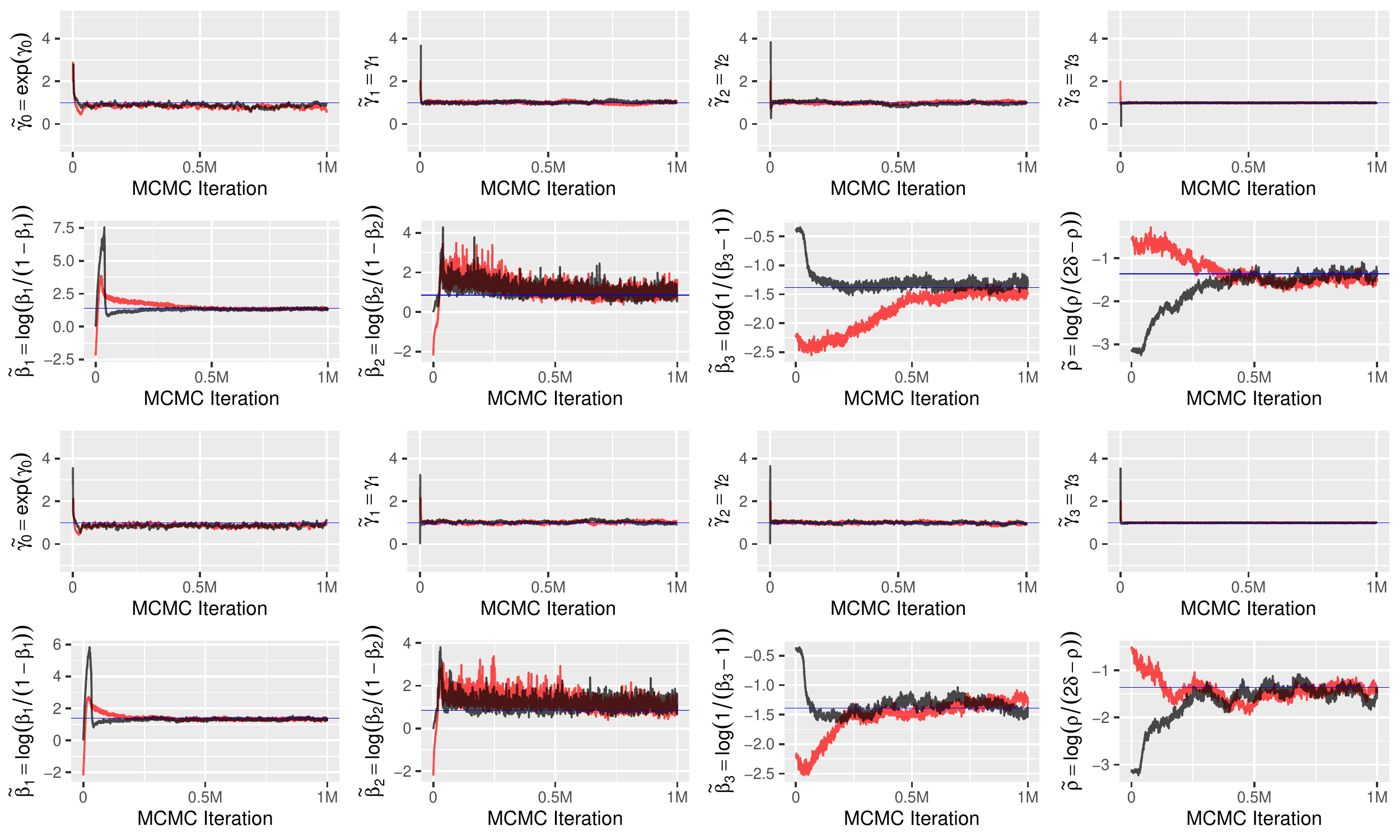}
\caption{Trace plots for all hyperparameters in our simulation study. The first two rows correspond to the Algorithm \ref{alg:Alg1} with batch size  $b =5$, and the last two rows are with batch size $b=20$. The two Markov chains (red and black), correspond to two different initial values. The total number of MCMC sample iterations is 1 million (1M). The blue horizontal lines show the true values.  }
\label{fig:TracePlotSim}
\end{figure}

Figure \ref{fig:TracePlotSim} shows the trace plots for all  hyperparameters for two different batch sizes $b$. The first two rows show the Markov chains for Algorithm \ref{alg:Alg1} with batch size $b=5$, and the last two rows correspond to batch size $b=20$. For both batch sizes, the MCMC chains show good and comparable mixing performance and relatively fast convergence, especially for the regression parameters $\gamma_0,\gamma_1, \ldots,\gamma_p$, as well as for the ``weights'' $\beta_1$ and $\beta_2$. The parameters $\beta_3$ and $\rho$ are slower to converge but they appear to suitably converge after about $0.5$ million iterations. The true values for all hyperparameters are close to the posterior means with a relatively narrow $95\%$ credible interval, suggesting that the MCMC algorithm performs well overall. The run-time is almost $9.5$ hours to run $1$ million iterations when the batch is size $b=5$, and it takes about $14$ hours to run $1$ million iterations when the batch size is set to $b=20$.  Significant speed-up can thus be obtained without compromising much on the convergence of Markov chains. Table \ref{tab:SGDSim} compares the performance of the SGLD-based MCMC algorithm (Algorithm \ref{alg:Alg1}) with respect to different batch sizes, $b=5,10$ and $20$. The results are similar for different batch sizes, so in our application we choose the batch size $b=5$ as it yields significant speed-up and the highest effective sample size per minute (ESS/min) for the hyperparameters that are the most tricky to estimate (i.e., $\beta_1$, $\beta_3$ and $\rho$).

\begin{table}[t!]
\small\addtolength{\tabcolsep}{2.5pt}
\centering
\caption{Absolute bias, standard error, $95\%$ credible interval (CI) length, effective sample size per minute (ESS/min.)  for the SGLD-based MCMC algorithm (Algorithm \ref{alg:Alg1}) for different batch sizes  $b=5, 10, 20$ in our simulation study. All posterior summary statistics are calculated after removing the first $3N_{tot}/4$ burn-in samples, where $N_{tot}=1$ million is the total number of MCMC iterations.}
\label{tab:SGDSim}
\begin{tabular}{c|c|c|c|c|c|c|c|c|c}

 &  batch size $b$ & $\exp(\gamma_0)$ & $\gamma_1$& $\gamma_2$ & $\gamma_3$ & $\beta_1$ & $\beta_2$ & $\beta_3$ & $\rho$\\
\hline
\multirow{3}{*}{Absolute  bias} & 5 & 0.18 & 0.01 & 0 & 0.01 & 0.01 & 0.02 & 0.08 & 0.03\\
& 10 & 0.09 & 0.02 & 0.04 & 0.01 & 0.01 & 0.07 & 0.13 & 0.04\\
& 20 & 0.13 & 0.03 & 0.03 & 0.01 & 0.01 & 0.05 & 0.24 & 0.03\\
\hline
\multirow{3}{*}{Standard error} & 5 & 0.06 & 0.03 & 0.03 & 0.01 & 0.01 & 0.03 & 0.16 & 0.02\\
& 10 & 0.05 & 0.03 & 0.04 & 0.01 & 0.01 & 0.03 & 0.22 & 0.03\\
 & 20 & 0.05 & 0.03 & 0.04 & 0.01 & 0.01 & 0.03 & 0.16 & 0.03\\
\hline
\multirow{3}{*}{$95 \%$ CI length} & 5 & 0.23 & 0.12 & 0.1 & 0.04 & 0.02 & 0.13 & 0.61 & 0.08\\
 & 10 & 0.18 & 0.12 & 0.13 & 0.03 & 0.02 & 0.12 & 0.83 & 0.12\\
 & 20 & 0.2 & 0.12 & 0.14 & 0.03 & 0.03 & 0.12 & 0.65 & 0.12\\
\hline
\multirow{3}{*}{ESS/min.} & 5 & 2.96 & 1.79 & 7.2 & 177.29 & 23.97 & 54.78 & 13.88 & 4.75\\
 & 10 & 9.66 & 1.69 & 2.25 & 150.67 & 16.43 & 57.35 & 2.41 & 2.23\\
 & 20 & 13.97 & 4.04 & 1.8 & 114.87 & 5.8 & 41.4 & 4.38 & 2.8\\
\end{tabular}
\end{table}

\begin{figure}[t!]
\includegraphics[width=1.0\textwidth]{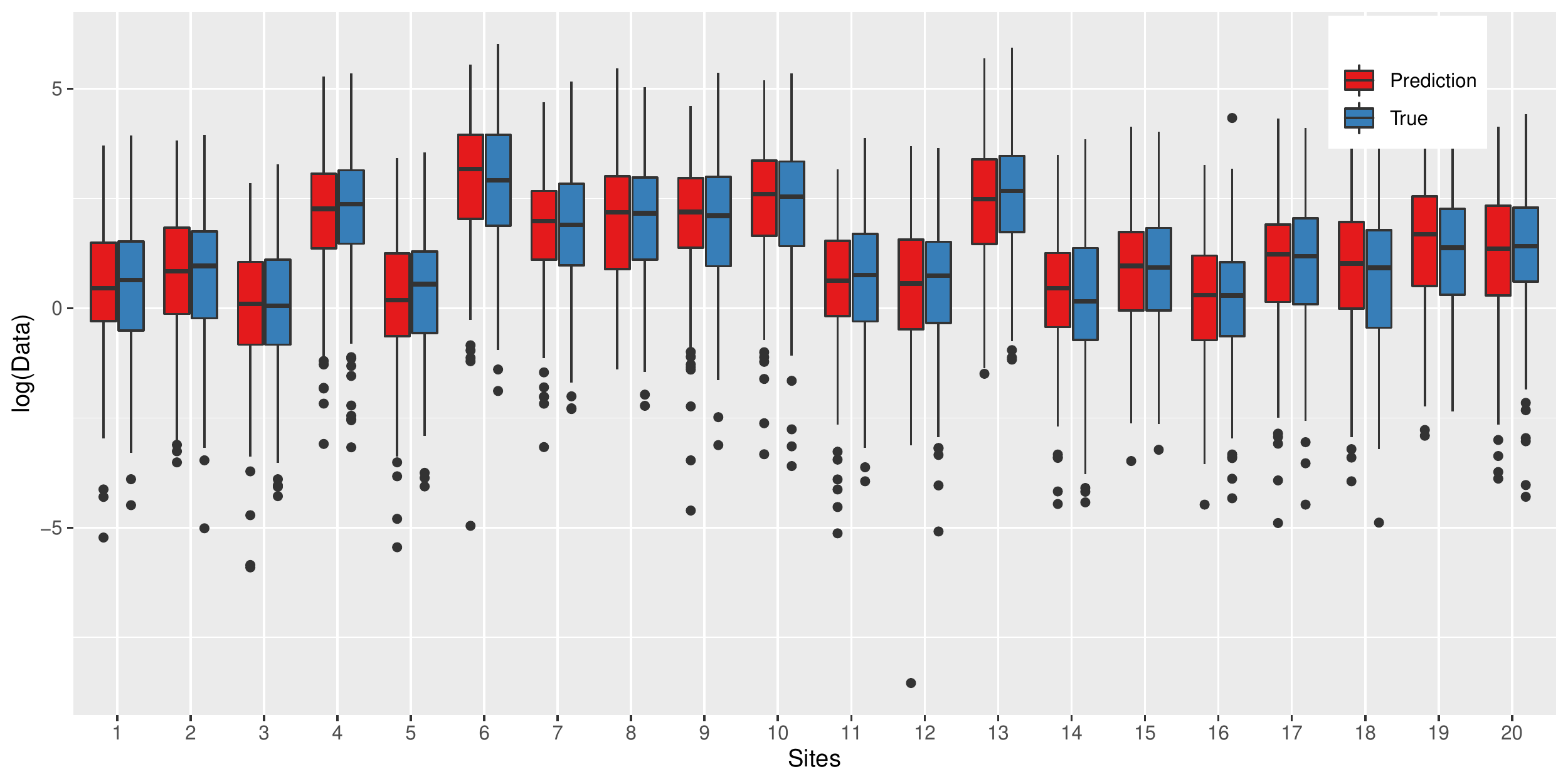}
\caption{Boxplots of  samples from the predictive distribution (red) at the 20 unobserved sites, along with boxplots of true (masked) observations at the same sites (blue), both in log scale. Here, the batch size of Algorithm \ref{alg:Alg1} is set to $b=5$. }
\label{fig: BoxPlotSim}
\end{figure}

 In Figure \ref{fig: BoxPlotSim}, we examine the predictive performance of our algorithm at the $20$ unobserved sites (with masked observations) by comparing boxplots of the true (masked) observations to samples from the corresponding posterior predictive distribution.  Samples from the posterior predictive distribution are obtained using the product mixture construction \eqref{eq:GenProdModel}, where the model hyperparameters are estimated using the sample posterior median. The posterior predictive boxplots are similar to the boxplots of the true data, and capture the variability at each site quite well. This suggests that our algorithm succeeds in performing spatial prediction.  Thanks to our model construction, it is possible to use our censored  inference approach to simultaneously fit the model to high threshold exceedances and perform spatial prediction at unobserved locations efficiently.  

\section{Application to extreme precipitation data in Spain}
\label{sec:application}
\subsection{Data description}
\label{subsec:DataDescription}
\begin{figure}[t!]
\begin{center}
\includegraphics[width=1\textwidth]{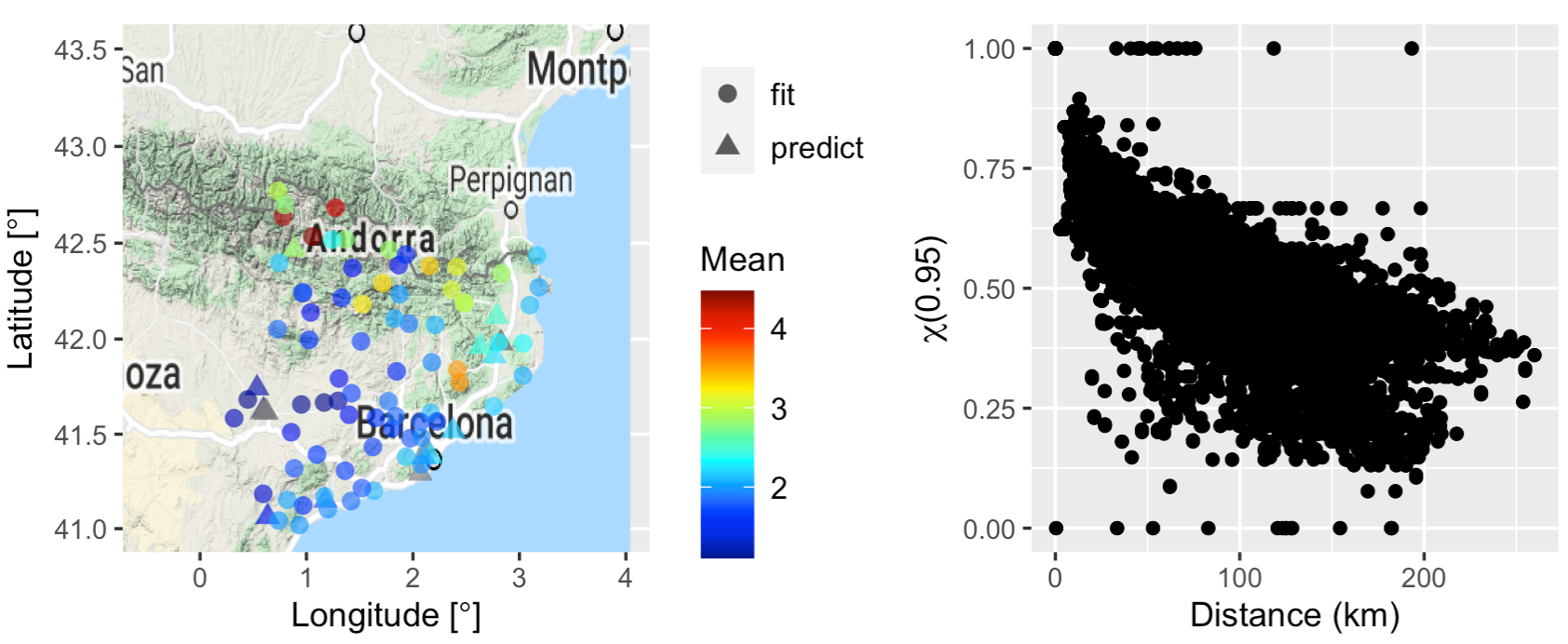}
\end{center}
\caption{Left: Color-coded mean precipitation amount [mm] at the study sites. Dots represent sites used for fitting and triangles represent sites where spatial prediction is performed. Right: $\chi(u)$ plot at a fixed threshold $u=0.95$. }
\label{fig:MapchiData}
\end{figure}

To illustrate our methodology, we now study precipitation intensities from Spain, publicly available from the \href{https://www.ecad.eu/}{European Climate Assessment and Dataset} project. The dataset reports daily precipitation amounts observed at more than $200$ spatial locations from $1941$ to $2018$. We apply our spatial product mixture model to a subset corresponding to a  study region in North-Eastern Spain with  $d=94$ observation sites, and we consider the study period  from $2011$ to $2020$. To avoid modeling complex temporal nonstationarites, we keep observations from September to December (i.e., the most rainy season), resulting in $n=1220$ temporal replicates (total $94\times 1220+ 1220=115,900$ latent variables). The distance between the two furthest sites is $260$km, and the two closest sites are $0.17$km apart.  The site-wise proportion of missing observations  varies from $0.01\%$ to $100\%$.  Also, the proportion of zeros at each site ranges from $41\%$ to $85\%$ with an average of $70\%$. The left panel in Figure \ref{fig:MapchiData} shows the site-wise mean precipitation plot, and the right panel shows the pairwise extremal correlation plot (i.e., $\chi(u)$ coefficient) at a fixed threshold $u=95\%$.  These plots show that there is considerably strong spatial heterogeneity and strong tail dependence in the data. The northern region receives higher precipitation than the southern region, and the sites at higher altitude receive higher precipitation, as well. This motivates  including geographical information such as latitude, longitude and altitude, as covariates in the model.

\subsection{Modeling precipitation intensities using the spatial product mixture models}
\label{subsec:DataApplModel}
We fit our specific spatial product model detailed in Section \ref{subsec:ExampleProdModel} to the Spanish precipitation data. Precisely, we consider several models where the scale vector $\bm\alpha$ comprises  linear or quadratic spatial covariates in terms of latitude $(\bm Z_1)$, longitude $(\bm Z_2)$, and altitude $(\bm Z_3)$, where covariates are standardized  to have mean zero and unit variance.  We  specifically consider the following  models:
 \begin{description}
 \small
  \item[M$1$:] $\boldsymbol{\alpha}=\exp(\gamma_0 \bm{1}_d)$
  \item[M$2$:] $\boldsymbol{\alpha}=\exp(\gamma_0 \bm{1}_d+\gamma_{\text{lat}} \textbf{Z}_1)$  
 \item[M$3$:] $\boldsymbol{\alpha}=\exp(\gamma_0 \bm{1}_d+\gamma_{\text{lat}} \textbf{Z}_1+\gamma_{\text{long}} \textbf{Z}_2)$ 
  \item[M$4$:] $\boldsymbol{\alpha}=\exp(\gamma_0 \textbf{1}_d+\gamma_{\text{lat}} \textbf{Z}_1+\gamma_{\text{long}} \textbf{Z}_2+\gamma_{\text{alt}}  \textbf{Z}_3)$  
 \item[M$5$:] $\boldsymbol{\alpha}=\exp(\gamma_0 \textbf{1}_d+\gamma_{\text{lat}} \textbf{Z}_1+\gamma_{\text{long}} \textbf{Z}_2+\gamma_{\text{alt}}  \textbf{Z}_3+ \gamma_{\text{lat}^2}  \textbf{Z}_1^2)$  
 \item[M$6$:] $\boldsymbol{\alpha}=\exp(\gamma_0 \textbf{1}_d+\gamma_{\text{lat}} \textbf{Z}_1+\gamma_{\text{long}} \textbf{Z}_2+\gamma_{\text{alt}}  \textbf{Z}_3+\gamma_{\text{lat}^2}  \textbf{Z}_1^2+\gamma_{\text{long}^2}  \textbf{Z}_2^2)$  
 \item[M$7$:] $\boldsymbol{\alpha}=\exp(\gamma_0 \textbf{1}_d+\gamma_{\text{lat}} \textbf{Z}_1+\gamma_{\text{long}} \textbf{Z}_2+\gamma_{\text{alt}}  \textbf{Z}_3+\gamma_{\text{lat}^2}  \textbf{Z}_1^2+\gamma_{\text{long}^2}  \textbf{Z}_2^2+\gamma_{\text{alt}^2}  \textbf{Z}_3^2)$  
  \item[M$8$:] $\boldsymbol{\alpha}=\exp(\gamma_0 \textbf{1}_d+\gamma_{\text{lat}} \textbf{Z}_1+\gamma_{\text{long}} \textbf{Z}_2+\gamma_{\text{alt}}  \textbf{Z}_3+\gamma_{\text{lat}^2}  \textbf{Z}_1^2+\gamma_{\text{long}^2}  \textbf{Z}_2^2+\gamma_{\text{alt}^2}  \textbf{Z}_3^2+\gamma_{\text{lat.long}}  \textbf{Z}_1 \textbf{Z}_2)$  
 \item[M$9$:] $\boldsymbol{\alpha}=\exp(\gamma_0 \textbf{1}_d+\gamma_{\text{lat}} \textbf{Z}_1+\gamma_{\text{long}} \textbf{Z}_2+\gamma_{\text{alt}}  \textbf{Z}_3+\gamma_{\text{lat}^2}  \textbf{Z}_1^2+\gamma_{\text{long}^2}  \textbf{Z}_2^2+\gamma_{\text{alt}^2}  \textbf{Z}_3^2+\gamma_{\text{lat.long}}  \textbf{Z}_1 \textbf{Z}_2+ \gamma_{\text{long.alt}}  \textbf{Z}_2 \textbf{Z}_3 )$  
 \item[M$10$:] $\boldsymbol{\alpha}=\exp(\gamma_0 \textbf{1}_d+\gamma_{\text{lat}} \textbf{Z}_1+\gamma_{\text{long}} \textbf{Z}_2+\gamma_{\text{alt}}  \textbf{Z}_3+\gamma_{\text{lat}^2}  \textbf{Z}_1^2+\gamma_{\text{long}^2}  \textbf{Z}_2^2+\gamma_{\text{alt}^2}  \textbf{Z}_3^2+\gamma_{\text{lat.long}}  \textbf{Z}_1 \textbf{Z}_2+ \gamma_{\text{long.alt}}  \textbf{Z}_2 \textbf{Z}_3 + \gamma_{\text{lat.alt}}  \textbf{Z}_1 \textbf{Z}_3)$  
  \end{description} 
 The censoring threshold for each of the models M1--M10 is set to the $75\%$  site-wise quantile of the strictly positive precipitation intensities. We choose $76$ stations for fitting and leave $18$ stations for prediction and model validation, where the true data are masked. For missing observations, we set the censoring threshold to $u_{tj}=\infty$.  For inference, we use the SGLD-based scheme (Algorithm \ref{alg:Alg1}) presented in Section \ref{sec:SimInferenceCens} with one million (1M) iterations, $N_m=25$, and with a batch size of $b=5$ for all the models. Also, we set $\delta_1=\delta_2=2$, to avoid numerical issue and given that the data shows strong dependence,  so that $\beta_1$ and $\beta_2$ parameters lies within the intervals  $[0,2]$ .
 
 Table \ref{tab:modelComp} compares the model performances based on the mean squared prediction error (MPE), the continuous ranked probability score  \cite[CRPS;][]{gneiting2007strictly}, and the tail-weighted CRPS \cite[twCRPS;][]{lerch2017forecaster}, averaged over the $18$ prediction stations.  Although model M3 is the best according to the CRPS and twCRPS criteria, it has surprisingly a huge MPE value. Moreover, quantile-quantile (QQ) plots for model M3 (not shown) reveal that the marginal fit is poor at several stations. Therefore, we prefer to select model M4 as our best model, which is slightly more complex but has the lowest MPE and comparable CRPS and twCRPS values, and can adequately capture the complex spatially-varying dynamics of precipitation extremes.  The total run-time for  model M4 is approximately 28 hours to generate one million samples. 

  \begin{table}[t!]
  \small\addtolength{\tabcolsep}{-1pt}
 \centering
 \caption{Mean squared prediction error (MPE) scores, continuous ranked probability score (CRPS), and tail-weighted CRPS scores, averaged over the  $18$  prediction stations, where the weight function in the twCRPS is chosen as  the Gaussian distribution function with variance $100$ and mean corresponding to the marginal threshold values. Lower values of (MPE), CRPS and twCRPS indicate better models, and the best performance for each criterion is highlighted in bold. }
 \label{tab:modelComp}
 \begin{tabular}{c|c|c|c|c|c|c|c|c|c|c}

 & M1 & M2 & M3 & M4 & M5 & M6 & M7 & M8 & M9 & M10\\

\hline
MPE & 3019.28 & 3046.96 & 3053.77 & \bf{2501.67} & 2926.22 & 2900.99 & 2869.43 & 2733.31 & 2730.02 & 2730.63\\

CRPS & 49.85 & 43.88 & \bf{43.75} & 45.06 & 43.94 & 44.05 & 44.06 & 47.47 & 47.38 & 48.27\\

twCRPS & 40.16 & 34.36 & \bf{34.25} & 35.35 & 34.36 & 34.45 & 34.45 & 37.62 & 37.55 & 38.36\\

\end{tabular}
 \end{table}

%
%
%
%
%
%
%
%
%

\begin{table}[t!]
  \centering
  \small\addtolength{\tabcolsep}{1pt}
 \caption{ Posterior summary statistics for our best model (M4), calculated based on $N_{tot}/4$ samples after deleting the first $3N_{tot}/4$  burn-in samples, where $N_{tot}=1$ million is the total number of MCMC samples.}
 \label{tab:SummaryTab}
\begin{tabular}{c|c|c|c|c|c|c|c|c}

& $\exp(\gamma_0)$ & $\gamma_{\text{lat}}$ &  $\gamma_{\text{long}}$ &  $\gamma_{\text{alt}}$ &  $\beta_1$ & $\beta_2$ &  $\xi=1/\beta_3$ &$\rho$ [km] \\

\hline

Post. mean & 3.247 & 0.089 & 0.116 & 0.246 & 1.258 & 1.995 & 0.095 & 519.406\\

Standard dev. & 0.062 & 0.023 & 0.027 & 0.019 & 0.011 & 0.005 & 0.002 & 0.046\\

Lower  $95\%$ CI & 3.137 & 0.04 & 0.064 & 0.208 & 1.237 & 1.981 & 0.092 & 519.313\\

Upper  $95\%$ CI	 & 3.31 & 0.13 & 0.163 & 0.279 & 1.288 & 2 & 0.089 & 519.337\\

ESS/min & 0.253 & 73.05 & 155.365 & 57.434 & 13.293 & 155.365 & 0.396 & 1.131\\

\end{tabular}
\end{table}

\begin{figure}[t!]
\includegraphics[width=1\textwidth]{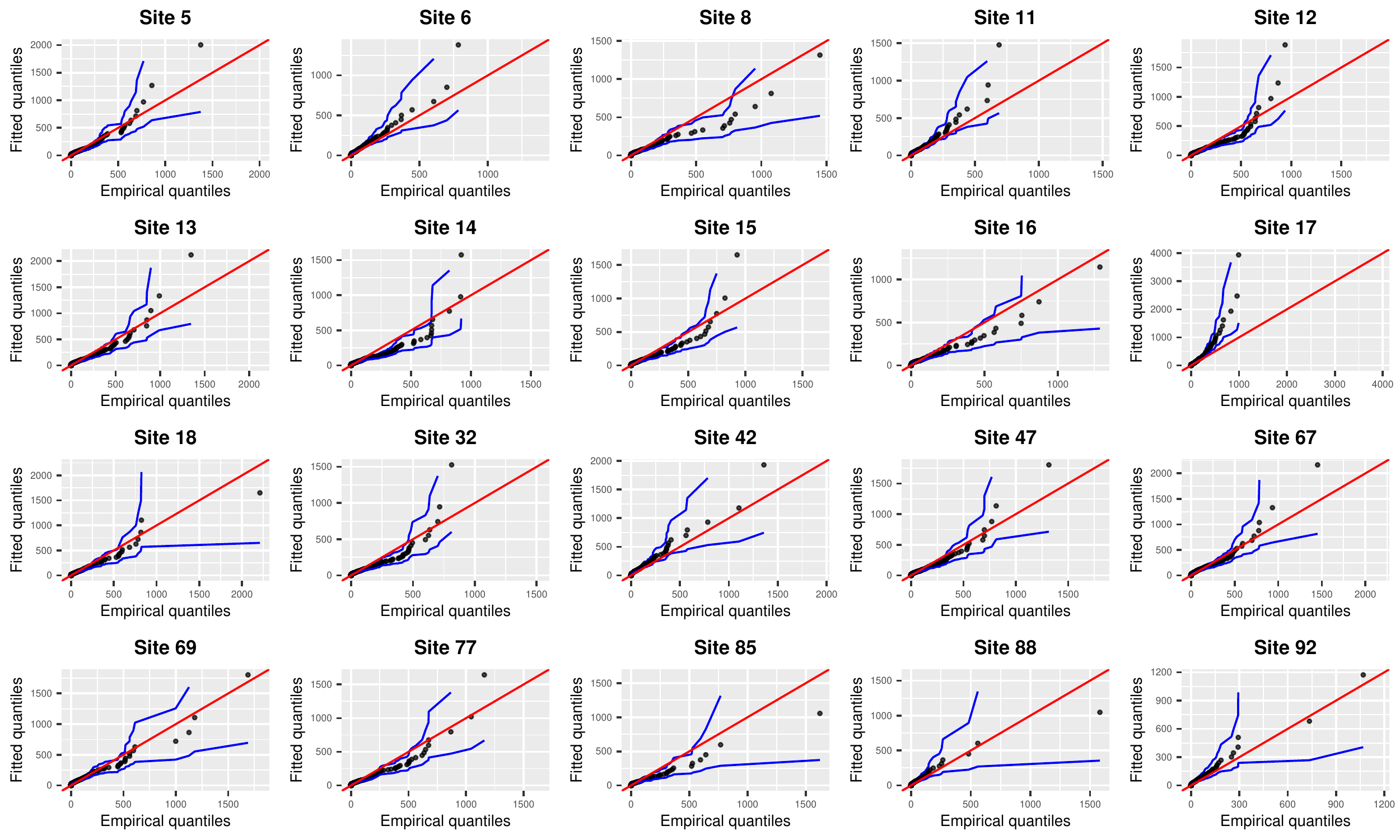}
\caption{QQ-plots at all prediction stations  where the data have been masked (up to site $\#18$), and some  selected fitted sites based on model M4. The fitted quantiles are simulated from the specific product model in Section \ref{subsec:ExampleProdModel}, where the model hyperparameters are estimated using the corresponding posterior mean,  calculated based on  $N_{tot}/4$ samples after removing the first $3N_{tot}/4$ burn-in samples, where $N_{tot}=1$M is the total number of MCMC samples.  }
\label{fig:QQplot}
\end{figure}
Table \ref{tab:SummaryTab} shows posterior  summary statistics for the best model M4.  The estimates of covariate  coefficients such as longitude ($\hat{\gamma}_{\text{long}}=0.12$), latitude ($\hat{\gamma}_{\text{lat}}=0.09$) , and altitude ($\hat{\gamma}_{\text{alt}}=0.25$), are highly significant, which shows that these  geographical covariates are needed in the model.  The estimate of $\beta_2$ is close to 2, which confirms that there is strong spatial dependence in the data and that the spatially constant term $\bm X_{2t}$  in \eqref{eq:GenProdModel} is crucially needed. The estimate of $\beta_1$  is about $1.26$, indicating that there is also non-negligible small-scale variability. The estimated tail index is $\hat{\xi}=0.10$, indicating moderately heavy tails, which is in line with most precipitation data.  The QQ-plots in Figure \ref{fig:QQplot}  show that the marginal predictive performance of our model is satisfactory. Figure \ref{fig:QQplot}  shows QQ-plots at all prediction stations where the data have been masked, and for some selected fitted stations.  The estimated quantiles in this plot are obtained by simulation from the fitted product mixture model of Section \ref{subsec:ExampleProdModel}, where the model hyperparameters are estimated using the posterior means of the corresponding posterior samples.																	
 \section{Conclusion}
\label{sec:conclusion}
In this paper, we have provided a constructive modeling framework for extreme spatial threshold exceedances based on product mixtures of three distinct and mutually independent random fields, where each of the fields is characterized by a distinct combination of heavy- or lighter-tailed margins and spatial dependence characteristics. 
These models provide high flexibility in the tail and at sub-asymptotic levels and may be used to capture strong tail dependence in high threshold exceedances. The dependence strength of our proposed model depends on the choice of the underlying copula structure at the latent level, and therefore we can get a variety of flexible dependence structures depending on the latent copula specification.

 We here design an SGLD-based MCMC algorithm to fit our model efficiently in high spatio-temporal dimensions, where the dimension of the latent parameter vector is comparable to the data dimension. By using the SGLD algorithm, we bypass the expensive calculation of full censored-likelihood, and full gradients, and hence inference is significantly faster with high data dimensions. Thanks to the SGLD algorithm, we can indeed drastically reduce the computational cost of each MCMC iteration, allowing us to fit complex Bayesian hierarchical models with strong data-level tail dependence, to threshold exceedances in  high dimensions. 
 
In our data application, we have shown how to model precipitation extremes, and have illustrated our methodology on data from North-Eastern Spain.  Although  our methodology was illustrated with the specific spatial mixture model of Section \ref{subsec:ExampleProdModel}, our constructive modeling framework is very general and can lead to several alternative heavy-tailed models. Future research directions include extending our spatial product mixture models to the spatio-temporal context. This may be performed by either incorporating temporal dependence in the fully spatially dependent latent parameter, $\bm X_{2t}$, or by introducing space-time dependence in the other latent parameter vector, $\bm X_{3t}$.  In the space-time context, products of more than three latent processes with distinct spatio-temporal characteristics may also be envisioned, and it would be interesting to explore how to further generalize our construction to capture different asymptotic regimes in space and time. 

\baselineskip 20pt
\newpage
\bibliographystyle{CUP}
\bibliography{ref}

\baselineskip 10pt

\end{document}